# Title

The macaque IT cortex but not current artificial vision networks encode object position in perceptually aligned coordinates.

**Abbreviated title:**

Perceptual position codes in IT

**Authors**

Elizaveta Yakubovskaya[1], Hamidreza Ramezanpour[1], Matteo Dunnhofer[1,2], and Kohitij Kar[1]

**Affiliation**

1. York University, Department of Biology and Centre for Vision Research, Toronto, Canada
2. University of Udine, Department of Mathematics, Computer Science, and Physics, Udine, Italy

\* Correspondence should be addressed to Kohitij Kar

E-mail: k0h1t1j@yorku.ca

**Conflict of interests**

The author declares no competing financial interests.

**Acknowledgments**

KK has been supported by funds from the Canada Research Chair Program (CRC-2021-00326), the Simons Foundation Autism Research Initiative (SFARI, 967073), Brain-Canada Foundation (2023-0259), the Canada First Research Excellence Funds (VISTA Program), and the National Sciences and Engineering Research Council of Canada (NSERC) Discovery Grant (RGPIN-2024-06223). EY received funding from the Connected Minds (supported by CFREF), the Queen Elizabeth II Graduate Scholarship in Science and Technology Program, and the NSERC CGSM. HR was supported by a CIHR Postdoctoral Fellowship. MD received funding from the European Union's Horizon Europe research and innovation programme under the Marie Skłodowska-Curie grant agreement n. 101151834 (CUP G23C24000910006).

# Abstract

Efficient interaction with the visual world requires not only accurate object identification but also precise localization of where objects are in space. While spatial ("where") processing has classically been attributed to dorsal stream pathways, recent work has shown that object position can also be decoded from ventral stream responses, including the inferior temporal (IT) cortex. However, because object position in these paradigms is tightly coupled to pixel-based location, it has remained unclear whether ventral stream position signals are perceptually meaningful or instead reflect incidental inheritance from retinotopic inputs. Here, we address this question by leveraging a classic visual illusion, the motion aftereffect, to dissociate perceived object position from retinal location while holding visual input constant. Combining large-scale intracortical recordings in macaque IT with matched human psychophysics, we show that motion adaptation induces systematic, direction-opponent biases in IT population codes for object position that qualitatively mirror human perceptual reports, despite unchanged pixel-level input. Motion adaptation reshapes IT representational geometry that likely contributes to these perceptual biases. Extending these findings to artificial vision systems, we observe that standard feedforward, recurrent, and state-of-the-art video-based neural networks fail to exhibit adaptation-induced position shifts, despite accurately encoding object position. Interestingly, imposing empirically derived IT-based transformations on model feature spaces is sufficient to simulate the effect, revealing adaptation-driven representational warping as a missing computational ingredient in current artificial vision systems. Together, these results identify IT as a locus of perceptually aligned spatial coding, reveal adaptation-driven representational restructuring as a mechanism linking neural dynamics to perception, and expose a principled gap between biological and artificial vision.

# Introduction

During daily tasks, humans extract and integrate a wide range of visual object properties. To perform even a simple task such as sorting apples from oranges, one must simultaneously know what the objects are and where they are located in space. The brain therefore needs to construct representations that combine information about both *what* an object is and *where* it is[1]. How and where these properties are encoded in the primate visual system remains a critical question in systems neuroscience.

A dominant view in the field, the two-stream hypothesis, posits a division of labor between cortical pathways: the ventral (what) pathway encodes object identity for perception, while the dorsal (where) pathway encodes spatial properties to guide action[2]. This account is supported by anatomical tracing studies in macaques that have revealed distinct projections from striate and prestriate areas to parietal versus inferior temporal (IT) cortex, giving rise to largely separate processing networks[3,4]. Lesion studies further support this view: patients and monkeys with temporal lobe damage show severe deficits in object recognition, whereas those with parietal damage struggle with visually guided reaching and grasping while retaining perceptual recognition abilities[5–8]. Over the past three decades, the two-stream hypothesis has generated important insights into the neural bases of visual cognition. Ventral stream areas have been established as critical substrates of high-level perceptual functions such as scene recognition, face and body processing, and object categorization[9–17]. Dorsal stream areas, in turn, have been shown to provide spatially precise codes that support object manipulation in task contexts[18–20], and have also been implicated in spatial navigation and motor imagery[21,22]. Yet despite its heuristic value, several lines of evidence challenge the idea that object identity and spatial properties are cleanly segregated between ventral and dorsal streams. Studies have demonstrated that ventral stream areas, including IT, encode not only category-level information but also properties such as size, rotation, and position[23,24]. Conversely, dorsal areas have been shown to encode category-level object information in contexts not directly tied to action[25]. Furthermore, there is growing evidence for extensive interactions between dorsal and ventral regions, particularly during tasks involving dynamic stimuli or high-level inference. Examples include perception of biological motion from point-light displays[26], object identification in kinematograms[27], and transformational apparent motion[28]. Even seemingly simple object-directed actions rely on reciprocal interactions between streams[29,30]. These findings suggest that the original dichotomy between "vision for perception" and "vision for action" needs to be revised in light of converging evidence for integrated processing.

Within this broader re-examination, the role of IT cortex has become especially central. Core object recognition studies have demonstrated that IT represents object identity in a linearly decodable format that directly predicts primate behavior[10,31,32]. Yet IT does not encode only identity. Recent work has shown that IT carries explicit information about category-orthogonal object properties such as size, position, and pose[23]. These findings suggest that IT representations extend beyond categorical coding and may serve as a hub for perceptually relevant object information, including spatial features typically attributed to the dorsal cortex. However, a key limitation of prior work is that object position in the test stimuli was always strongly correlated with pixel-based ground-truth position. As a result, decoders trained on IT responses could successfully recover object position, but it remained unclear whether this success reflected a genuinely perceptual code or a trivial by-product of feedforward retinotopic pooling (see[33] for a review). To dissociate these explanations, stronger tests are needed in which perceived position diverges from pixel-based position. Importantly, modern artificial neural network (ANN) models of the ventral stream, that approximate IT responses[34] and recognition[35], behavior, exhibit similar position- and pose-related signals[23], raising the question of whether these models capture the same perceptually meaningful spatial representations as biological IT, or whether both systems merely reflect inherited properties of their visual inputs.

In this study, we sought to provide such a test by leveraging the motion aftereffect (MAE). In the MAE paradigm, prolonged exposure to motion in one direction biases the perceived position of subsequently viewed stationary stimuli, which appear shifted opposite to the direction of adaptation[36]. Critically, because the retinal input and pixel-level object location remain unchanged, the MAE provides a strong null prediction: any system whose position signals are purely feedforward or statically inherited from retinotopic inputs should exhibit no change in encoded object position following adaptation. The neural basis of the MAE has been most closely linked to adaptation in the middle temporal (MT) area, where direction-tuned neurons undergo repetition suppression after prolonged stimulation[37–40]. However, motion adaptation effects have also been observed in ventral areas. V4 contains motion-direction selective domains whose responses are modulated by adaptation[41–44]. IT neurons respond not only to static object features but also to both simple and complex motion stimuli, ranging from gratings and dot patterns to bodies and objects in motion[45–49]. Critically, disrupting IT function impairs motion discrimination, underscoring the behavioral relevance of these signals[49]. These findings raise the possibility that motion adaptation may systematically reshape IT representations in ways that link neural signals to perceptual reports. Indeed, IT is known to exhibit robust adaptation phenomena, including repetition suppression to repeated shapes or objects[6,50]. Adaptation not only reduces firing rates but can also sharpen representational distinctions, improving classification of stimuli that differ from the adapter[51]. By analogy, one might expect motion adaptation to modulate IT object codes in a manner that produces systematic perceptual biases in position. On the other hand, in the absence of such adaptive representational restructuring, as expected under a static encoding model or in standard ANNs, no position shift should be observed.

Finally, we extended this line of questioning into the computational domain by asking whether current ANN models of the ventral stream provide suitable models of these dynamic processes. Feedforward ANNs and their modern variants have been shown to approximate IT object codes and human behavior for core object recognition[23,52]. More recently, models with recurrent or temporal dynamics have been proposed as closer analogs of biological vision[53,54]. Importantly, explicit implementation of repetition suppression in ANNs has been shown to replicate several behavioral and neurophysiological adaptation effects[55]. We therefore asked whether motion adaptation-like transformations in IT could be reproduced in-silico. Specifically, we (i) quantified response suppression in IT neurons, (ii) implemented analogous decay functions in object recognition networks, and (iii) probed temporal dynamics in video ANNs. If adaptation-induced position biases were to emerge naturally in such models, it would suggest that existing ANN architectures capture some of the key mechanisms underlying IT's dynamic perceptual alignment.

Conversely, their failure to reproduce such biases would point to critical gaps in current modeling approaches.

In summary, this study addresses a longstanding gap between neural evidence for object position signals in IT and the question of whether these signals are perceptually meaningful. By leveraging motion adaptation, we created conditions where perceived and pixel-based positions diverge, enabling us to test the behavioral relevance of IT codes. We show that IT population responses shift in ways that align with human perceptual reports, identify a subpopulation of neurons whose adaptation most strongly contributes to this effect, and demonstrate that standard ANN models do not yet capture the underlying dynamics. These findings advance our understanding of IT as a locus where spatial information is encoded in perceptually aligned coordinates and provide a new benchmark for evaluating models of dynamic vision.

# Results

As outlined above, we sought to test whether object position information in the macaque IT cortex reflects perceptually relevant signals. To do so, we combined intracortical recordings in macaques, human psychophysics, and in-silico modeling. Our approach was organized in stages: first, we established a baseline by replicating prior evidence that IT activity contains explicit information about object position[23]. We then examined how motion adaptation alters the representation of images across the IT cortex and whether these changes qualitatively align with human perceptual biases for localizing objects. Finally, we evaluated whether artificial neural networks (ANNs) can reproduce these adaptation-induced representational shifts.

**Object positions can be reliably reported by human observers and is explicitly represented in the macaque IT cortex**

We first asked whether object position can be measured reliably at the level of human perception in our image-set (examples shown in Figure 1A). Images from eight object identities (bear, elephant, face, car, dog, apple, chair, and airplane) were embedded in naturalistic backgrounds at systematically varied positions across the image. To establish a behavioral benchmark (see Methods for more details), we presented a subset (40 images) of these images briefly (100 ms within the central 8 deg) to human participants (n = 35). Following the image presentation, participants performed a behavioral localization task by making a mouse click on a blank frame (spanning the central 8 deg) indicating the perceived centroid of the objects (Figure 1B). Human responses were highly reliable ($r_x$ = 0.96, $r_y$ = 0.99) across repetitions (i.e. unique observers in our case), providing a precise estimate of perceptual object position.

To validate the presence of behaviorally relevant object position codes in macaque IT, we began by recording large-scale population responses during passive viewing of naturalistic object images. We analyzed IT responses from chronically implanted electrode arrays and trained cross-validated linear regression models to predict horizontal (x) and vertical (y) object positions from population vectors derived from spike counts measured 70–170 ms after image onset—a temporal window previously shown to capture behaviorally aligned IT activity for both object classification[31] and object position[23]. Consistent with prior findings, object position could be robustly decoded from IT responses (Figure 1C). Prediction accuracy increased with the number of simultaneously recorded neurons, reaching correlations of r ≈ 0.60 for x-position and r ≈ 0.68 for y-position when >150 units were included, closely replicating previous results[23].

To place these neural results in context, we applied the same decoding framework to feature activations from several widely used object recognition ANNs (Figure 1D). Position information was recoverable from some architectures such as VGG-16 and ResNet-18 (VGG-16 $r_x$ = 0.84; ResNet-18 $r_x$ = 0.81 ; VGG-16 $r_y$ = 0.87, ResNet-18 $r_y$ = 0.85. Other architectures, including transformer-based models (ViT-L32), showed markedly weaker position decodability overall ($r_x$ =

0.59, $r_y$ = 0.70). These results confirm that while object position information is broadly present in visual systems, IT cortex carries distinct and behaviorally relevant position codes that align with human perceptual estimates.

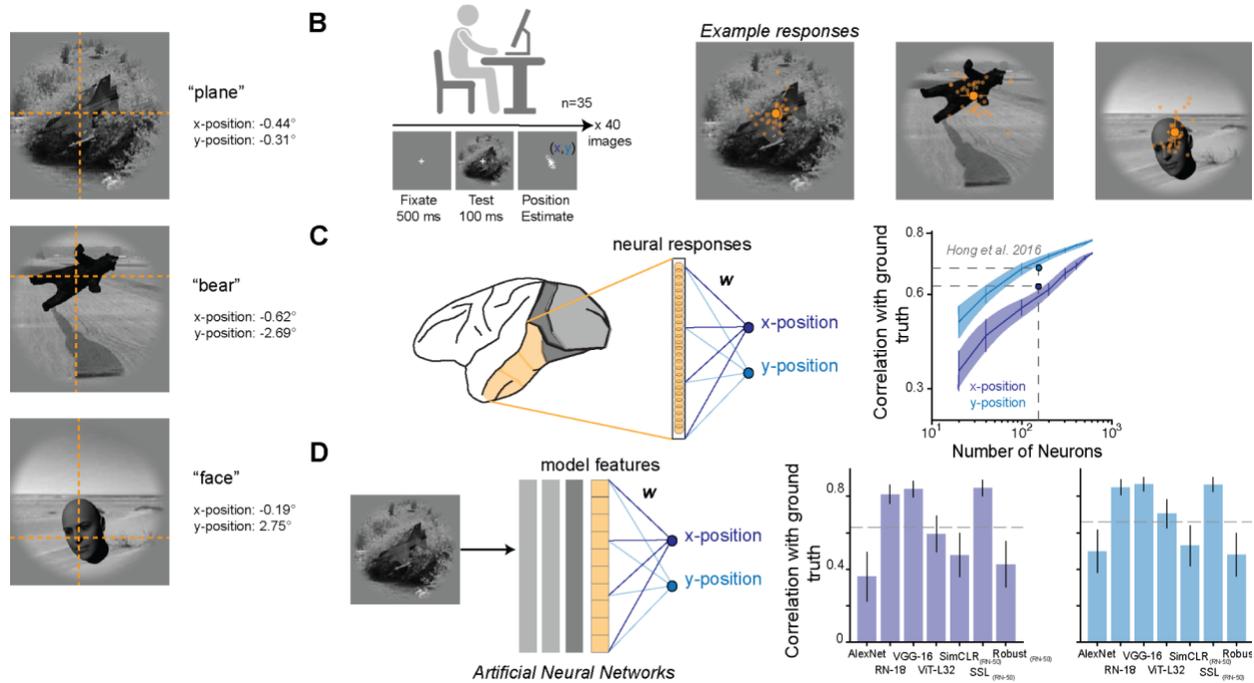

Figure 1. Object position information is explicitly represented in macaque IT cortex. (A) Example stimuli from the synthetic naturalistic image set. Objects (plane, bear, face) were embedded in textured backgrounds and presented at systematically varied positions. Crosshairs denote the pixel-based ground-truth object centers (x, y), reported in units of visual degrees (°) from the centre of the image at 0° along the x and y-axes. Positive values along the x-axis denote objects to the right of the centre. Positive values along the y-axis denote objects below the centre. (B) Human behavioral localization task. Participants (n = 35) fixated a central cross, viewed a test image (100 ms), and then reported the perceived object center. Example responses (orange dots) are shown for three images, illustrating variability across trials. (C) Neural decoding of object position. Schematic of experimental setup and decoding framework: macaque IT responses during image viewing were used to train linear decoders for x- and y-positions. Right, decoder performance increased with the number of neurons included, reproducing Hong et al. (2016), with correlations plateauing around r ≈ 0.6–0.7. Shaded regions indicate SEM across bootstrap iterations. (D) Object position decoding from ANN features. Linear decoders were trained on features from IT-aligned layers of different object recognition models. Bar plots show mean correlation with ground-truth object positions (± 95% confidence intervals) across models. Position information was broadly recoverable. Dashed lines mark IT benchmark[23] correlations ($r_x$ = 0.63, $r_y$ = 0.66).

Together, these results replicate the finding that IT cortex explicitly encodes object position and extend it by showing how these representations compare to those recoverable from leading ANN models. However, because ground-truth pixel positions and perceived positions are aligned in these paradigms, these results cannot distinguish whether IT signals are perceptually meaningful or simply reflect feedforward pooling of retinotopic receptive fields. To resolve this ambiguity, we next sought to create a scenario where perceived object position diverges from its pixel-based location, enabling us to test the perceptual alignment of IT codes.

**Motion adaptation reliably induces position biases in human observers**

We asked whether prolonged motion adaptation could produce reliable shifts in human position judgments (Figure 2A). Participants (n = 22) fixated centrally while viewing adapter stimuli consisting of drifting gratings (rightward or leftward, 30 s initial adaptation followed by 3 s top-up periods). After adaptation, similar to the previous object localization task, subjects were briefly shown a stationary object image (100 ms) and then reported its perceived center on a blank screen via mouse click.

As expected, motion adaptation induced systematic lateral biases in perceived position (Figure 2B). Following rightward motion adaptation, object centers were perceived as shifted leftward relative to estimates before motion adaptation; conversely, after leftward motion adaptation, centers were shifted rightward. These shifts were evident at the group level as a reliable deviation from the unity line in a scatterplot of mean x-position estimates under leftward and rightward adaptation (Figure 2B). An example image illustrates how the same object was localized differently depending on adapter direction (Figure 2B).

Group-level quantitative analysis confirmed these biases (Figure 2C). For both adapter conditions, x-position estimates were displaced in the direction opposite to the adapter motion (rightward adapter: leftward shift, Mean $\Delta P_x$ = -0.20°, SD = 0.32°, t(39) = -3.91, $p < 0.001$, single-sample t-test; leftward adapter: rightward shift, Mean $\Delta P_x$ = 0.13°, SD = 0.26°, t(39) = 3.10, p = 0.0036, single-sample t-test). By contrast, no systematic changes were observed for y-position estimates (rightward adapter: Mean $\Delta P_y$ = 0.031°, SD = 0.38°, t(39) = 0.50, p = 0.62, single-sample t-test; leftward adapter: Mean $\Delta P_y$ = -0.0095°, SD = 0.32°, t(39) = -0.18, p = 0.85, single-sample t-test), consistent with the selectivity of the manipulation to horizontal motion.

To ensure that these perceptual shifts reflected genuine adaptation effects rather than variability across sessions, we assessed the reliability of human position estimates across conditions (Figure 2D). Position estimates were highly correlated across repetition (before and after rightward motion adaptation: $r_x$ = 0.92, $r_y$ = 0.97; before and after leftward motion adaptation: $r_x$ = 0.95, $r_y$ = 0.98; after rightward and leftward motion adaptation: $r_x$ = 0.96, $r_y$ = 0.98), with no evidence of degradation in internal consistency of estimates following adaptation (before motion adaptation: $r_x$ = 0.96, $r_y$ = 0.99; after rightward motion adaptation: $r_x$ = 0.93, $r_y$ = 0.97; after leftward motion adaptation: $r_x$ = 0.96, $r_y$ = 0.98). This indicates that motion adaptation systematically biases perceptual localization without increasing noise in estimates.

Together, these results confirm that motion adaptation provides a robust behavioral assay in which perceived and pixel-based object positions are dissociated. Having confirmed that motion adaptation produces robust lateral shifts in human position judgments (Figure 2), we next asked whether macaque IT population responses undergo corresponding changes that could underlie these percepts. Specifically, we examined (i) how adaptation modulates IT firing and population geometry, and (ii) whether decoders trained on pre-adaptation IT activity express the same direction-opponent positional biases when applied to post-adaptation responses.

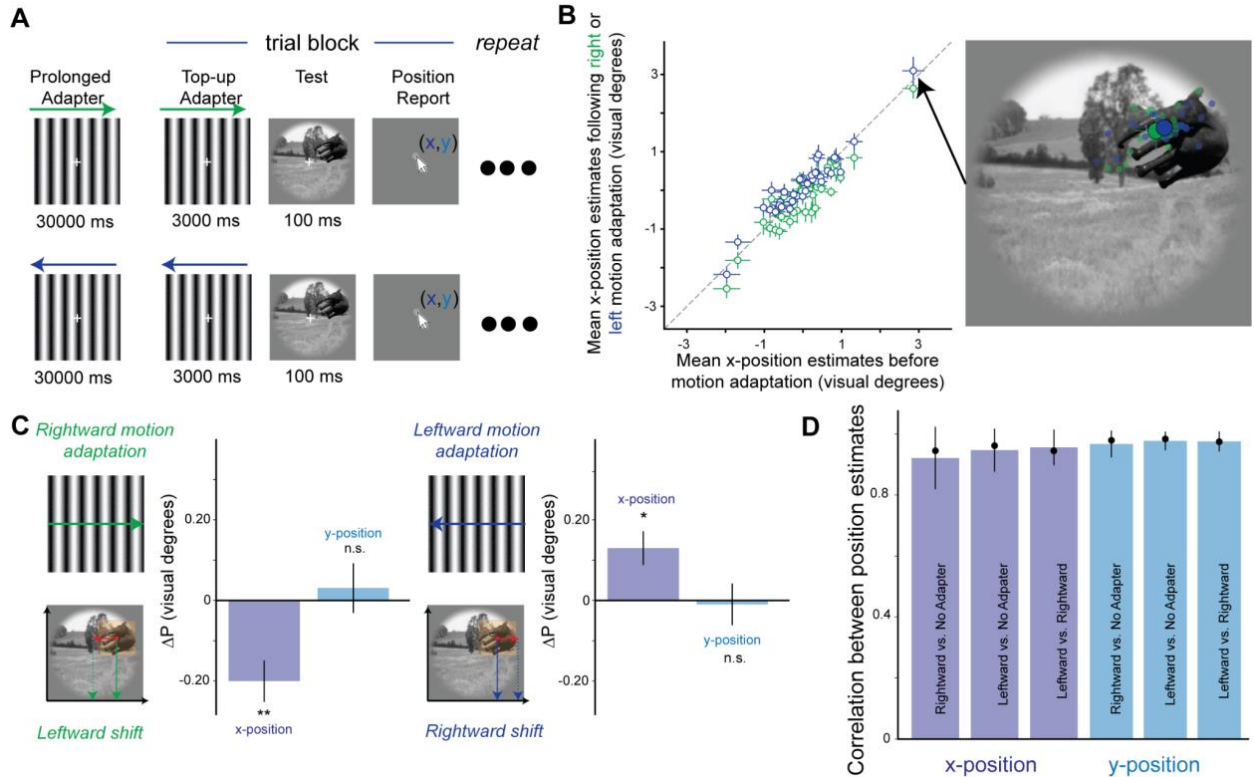

Figure 2. Motion adaptation reliably induces systematic position biases in human observers. (A) Human psychophysics paradigm. Participants were adapted to prolonged drifting gratings (rightward or leftward, 30 s initial exposure, followed by 3 s top-up periods) before briefly viewing a stationary test image (100 ms) and reporting its perceived center. (B) Scatterplot of mean x-position estimates across images following leftward versus rightward motion adaptation, reported in units of visual degrees (°) from the center of the image at 0°, along the x-axis. Points above the unity line indicate rightward perceptual shifts after leftward adaptation (blue) and points below the unity line indicate leftward shifts after rightward adaptation (green). Error bars denote standard error of the mean across trials. Inset shows an example test image with overlaid individual responses (green: right-adapted, blue: left-adapted), demonstrating systematic lateral displacement. Error bars denote standard error of the mean across trials. (C) Group-level position biases. Leftward shifts in estimates were observed after rightward adaptation, and rightward shifts after leftward adaptation. No significant changes were observed along the vertical (y) axis. Error bars denote standard error of the mean; **p < 0.001, *p < 0.05. (D) Reliability of position estimates. Correlations of reported positions across adaptation and control conditions (± 95% confidence intervals) remained high for both x- and y-dimensions, confirming that adaptation altered perceived position without degrading estimate reliability. Black dots indicate correlation ceiling, with error bars denoting standard deviation across repetitions of split-half reliability estimates.

## Motion adaptation suppresses firing and shifts IT population geometry

During adapter presentation (drifting gratings), mean IT responses decayed over time, consistent with repetition-suppression–like dynamics. To quantify unit-level stability, we correlated each neuron's response vector to the image set before vs. after adaptation (separately for leftward and rightward adapters). As expected, these correlations were broadly distributed across the population, indicating heterogeneous sensitivity to adaptation (Figure 3B, center histograms; example neurons on either side). Notably, examining the difference between the top and bottom histograms suggests that individual units are not necessarily equally sensitive to adaptation to both directions of motion. To assess how adaptation altered neural response patterns, we compared each neuron's image-response profile before and after adaptation. Individual units showed heterogeneous effects: some neurons exhibited substantial changes in their response vectors across images, whereas others remained relatively stable (Figure 3B). This variability indicates that motion adaptation does not simply scale responses uniformly across neurons but can reshape the pattern of activity across the population. We next quantified how these changes impacted the structure of the population representation using centered kernel alignment (CKA[56]), which measures similarity between high-dimensional neural response spaces. To interpret these similarities, we estimated a reliability ceiling for each condition using split-half analyses: trials were randomly divided into two halves, CKA was computed between the resulting response matrices, and the values were corrected using the Spearman–Brown formula. The geometric mean of the two split-half reliabilities provided an estimate of the maximum representational similarity expected given the noise in each dataset (Methods). The similarity between baseline responses and responses following motion adaptation was markedly reduced for both rightward (mean CKA: 0.36, ceiling: 0.56), and leftward (mean CKA: 0.32, ceiling: 0.37) adapters (Figure 3C).

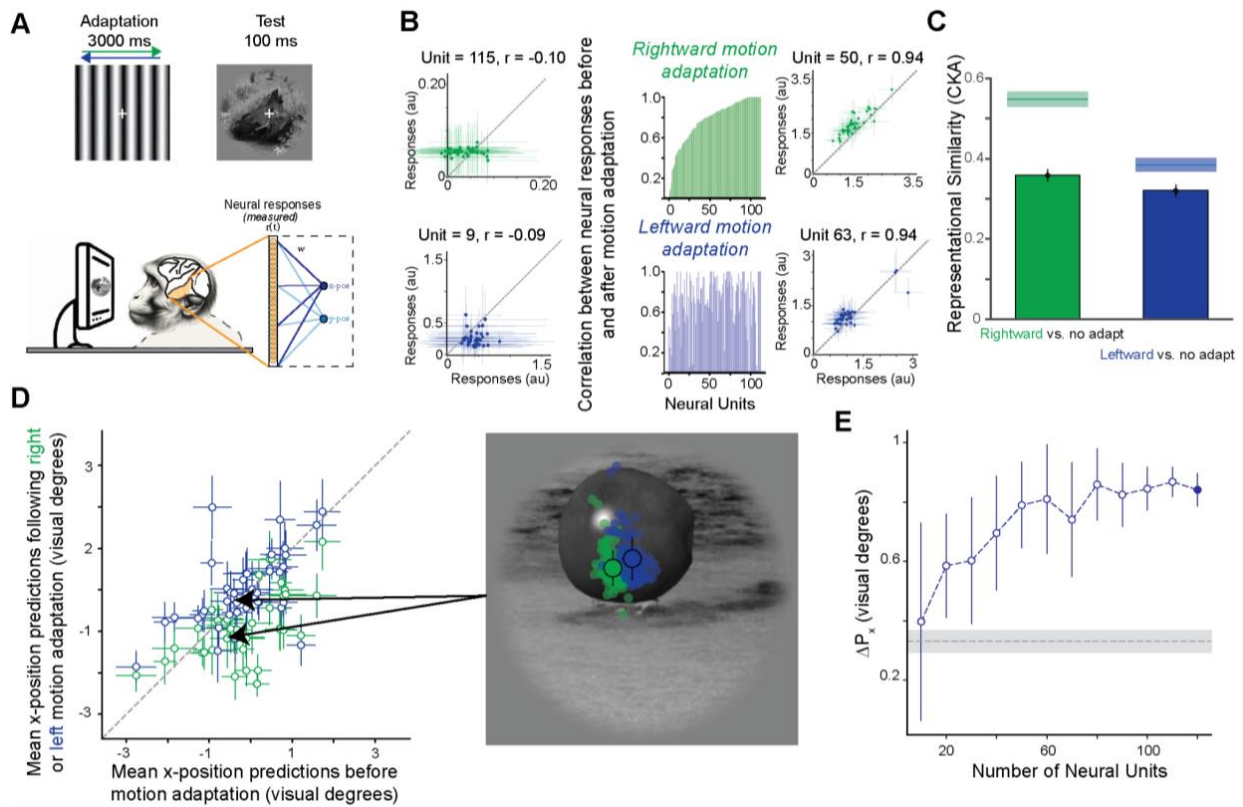
121212

**Figure 3. Motion adaptation reliably induces systematic changes in macaque IT representations of images and biases object position predictions. (A)** Experimental paradigm and decoding framework. Monkeys viewed drifting gratings for 3 s to induce motion adaptation (leftward or rightward), followed by brief presentation of test images (100 ms). Neural responses in inferior temporal (IT) cortex were recorded and used to decode object position from the population activity pattern. **(B)** Correlation between neuronal responses before and after motion adaptation (green: right-adapted, blue: left-adapted). Errors denote standard deviation across repetitions of object position prediction with different sets of train-test splits. Scatter plots on either side of the center bar graphs represent individual, exemplar units with weak or strong correlations between responses before and after motion adaptation, with error denoting standard deviation. Units differ in sensitivity to either direction of motion adaptation. **(C)** Motion adaptation alters IT representational geometry. Representational similarity between neural response spaces was quantified using centered kernel alignment (CKA). Bars show similarity between baseline responses and responses following rightward or leftward adaptation. Shaded regions indicate reliability ceilings estimated from split-half analyses. Both adaptation conditions show reduced similarity relative to baseline and fall below their respective ceilings, indicating a genuine shift in the geometry of population responses rather than noise alone. Error bars denote median absolute deviation. **(D)** Left: Scatterplot of mean x-position estimates across images following leftward versus rightward motion adaptation. Points above the unity line indicate rightward perceptual shifts after leftward adaptation (blue) and points below the unity line indicate leftward shifts after rightward adaptation (green). Right: Inset shows an example test image with overlaid individual responses (green: right-adapted, blue: left-adapted), demonstrating systematic lateral displacement of predictions. For both Left and Right, error bars denote standard deviation across repetitions of object position prediction with different train-test sets. **(E)** Group-level biases ($\Delta P_x = P_{x, \text{adapt left}} - P_{x, \text{adapt right}}$) in x-position predictions as a function of the number of neural units contributing to predictions. The final, colored point denotes the same statistic from the whole population of neurons. Error bars denote standard deviation across subsamples. The grey dashed line indicates group-level biases computed from human position estimates and the shaded region denotes standard error of the mean across trials. Biases in x-position predictions from macaque IT have the same directionality (i.e., opposite the adapter direction) as those in human position estimates.

## IT position decodes show direction-opponent biases after adaptation

We then asked whether these representational geometry changes carry perceptual consequences. L2-regularized linear decoders were trained on IT responses to the stationary images (70–170 ms after onset), without prior adaptation, and then applied, without retraining, to responses evoked by the same images following rightward or leftward adaptation. Decoded x-positions were systematically displaced opposite the adapter direction (Figure 3D): after rightward adaptation, estimates shifted leftward; after leftward adaptation, estimates shifted rightward. At the group level, these effects were significant only following rightward motion adaptation (leftward shift, Mean $\Delta P_x$ = -0.71°, SD = 0.80°, t(39) = -5.52, p < 0.001, single-sample t-test). Following leftward motion adaptation, position predictions showed a non-significant trend towards rightward shift (Mean $\Delta P_x$ = 0.14°, SD = 0.85°, t(39) = 0.98, p = 0.33, single-sample t-test). Importantly, the difference between position estimates following leftward and rightward motion adaptation ($\Delta P_x = P_{x, \text{adapt left}} - P_{x, \text{adapt right}}$) had the same directionality as that was observed in human estimates (Figure 3E, Supplementary Figure 1), suggesting that the effects of motion adaptation on object position representations in macaque IT are comparable to those on human object position estimation. Furthermore, y-position decodes showed no systematic change (rightward adapter: Mean $\Delta P_y$ = 0.16°, SD = 0.95°, t(39) = 1.03, p = 0.31, single-sample t-test; leftward adapter: Mean $\Delta P_y$ = -0.055°, SD = 1.04°, t(39) = -0.33, p = 0.74, single-sample t-test), as expected for a horizontal manipulation. Importantly, the reliability of decoded positions before ($r_x$ = 0.81, $r_y$ = 0.83) and after motion adaptation (right-adapted $r_x$ = 0.67, $r_y$ = 0.73 ; left-adapted $r_x$ = 0.77, $r_y$ = 0.75) remained high, indicating that adaptation produced a *bias* rather than an increase in noise. Together, these results show that motion adaptation both (i) alters IT population codes and (ii) imposes behavior-like shifts on position readouts from those codes, providing strong evidence that IT represents object position in perceptually aligned coordinates rather than as a trivial by-product of feedforward retinotopy

So far our results show that motion adaptation systematically alters IT population geometry and biases position decodes in a manner that mirrors human perceptual reports. However, while these findings establish IT as a potential, perceptually aligned substrate for object position, they also raise a critical computational question: can existing ventral stream ANN computations capture these dynamics? Standard feedforward models should, by design, predict identical object positions before and after adaptation, since the pixel-based input remains unchanged. Such invariance highlights a key shortcoming of current architectures: they lack the intrinsic neuronal dynamics that generate perceptual biases from identical stimuli. To probe this issue, we next asked whether imposing neuron-like transformations on ANN features, derived directly from IT adaptation data, would be sufficient to recapitulate the observed position shifts.

**Inducing neuron-like transformations in ANN features produces IT-like positional biases**

Feedforward object recognition ANNs, when probed in their standard form, make a straightforward prediction: object position read out from model features should remain unchanged before and after adaptation, because the pixel-based location of the object is identical across conditions. In other words, such models cannot account for the dissociation between retinal input and perceptual experience that characterizes the motion aftereffect. This highlights a fundamental limitation: unlike biological neural populations, current feedforward ANNs lack the intrinsic dynamics that could induce systematic shifts in representational geometry. To ask whether neuron-like adaptation could, in principle, drive ANN features toward perceptually aligned states, we implemented a cross-condition linear transformation framework (Figure 4A). Using IT population responses, we derived linear mappings that transformed baseline activity into its adapted counterpart for each motion direction. These empirical transformations were then applied in reverse to ANN features, a process we term "neuralization", to approximate how adaptation might reshape their representational space. Remarkably, once neuralized, ANN features expressed the same kind of position biases observed in macaque IT and human observers (Figure 4B). Across diverse architectures (VGG-16, ResNet-18, SimCLR), decoders trained on baseline features predicted systematically different object positions after applying the leftward versus rightward neuralization transform. These displacements were direction-opponent and resembled the behavioral and neural biases reported above. Thus, while unmodified ANNs fail to capture the consequences of motion adaptation, imposing empirically derived, neuron-like transformations is sufficient to recapitulate the effect. This result demonstrates that the perceptually aligned position biases we observed in IT are computationally achievable within current ANN feature spaces, but require dynamics beyond those implemented in standard feedforward models.

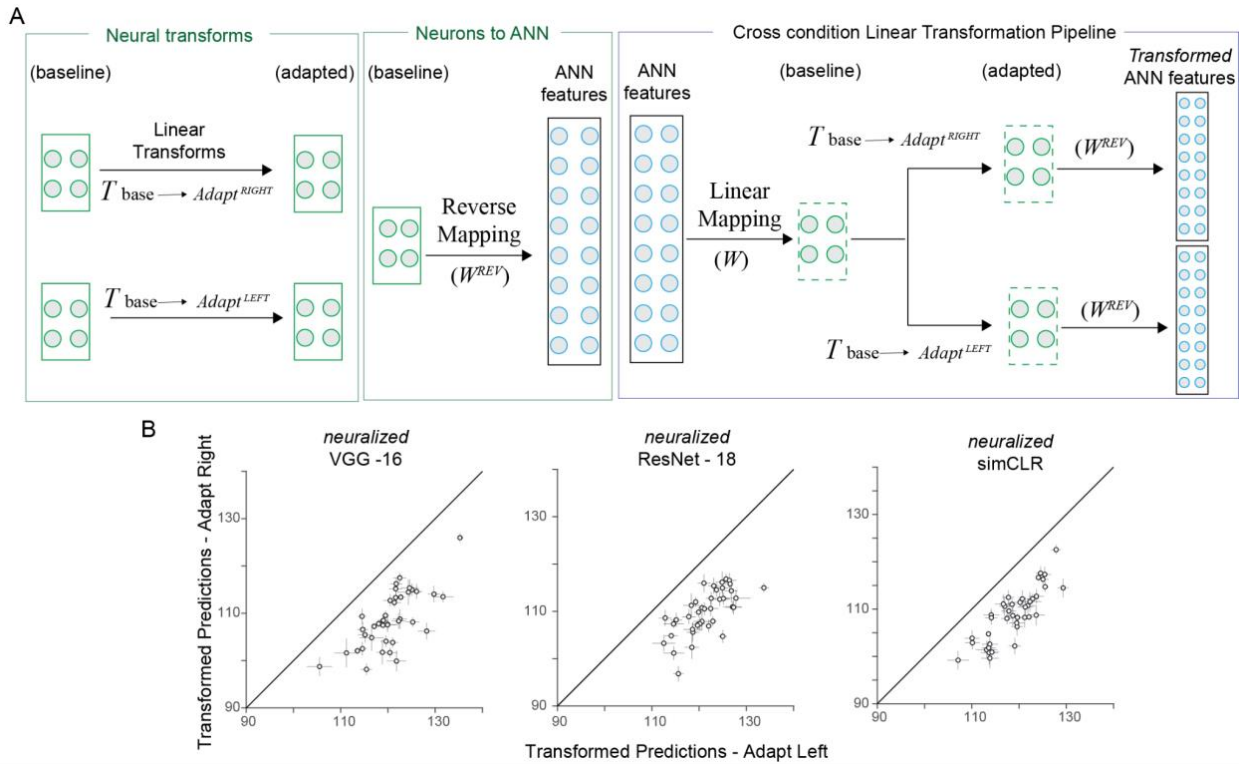

**Figure 4. Neuralizing ANN features with IT-derived transformations induces position biases. (A)** Schematic of the transformation framework. Left, IT baseline responses were linearly transformed into their adapted counterparts (for rightward or leftward motion). Middle, reverse mappings were used to project between IT activity and ANN feature spaces (similar to[57]). Right, applying these mappings to baseline ANN features yielded "neuralized" ANN responses that approximate how adaptation reshapes representational geometry. **(B)** Position decodes from neuralized ANN features. Scatterplots show mean predicted positions after leftward versus rightward neuralization for VGG-16, ResNet-18, and SimCLR. Points deviating from the unity line reflect direction-opponent position biases, with rightward neuralization producing leftward shifts and leftward neuralization producing rightward shifts. These systematic displacements mirror the biases observed in macaque IT and human behavior.

## Testing whether intrinsic ANN dynamics can account for adaptation-induced biases

The neuralization experiments (Figure 4) demonstrated that imposing empirically derived IT-like transformations on model features is sufficient to induce perceptually aligned positional biases. However, this approach is externally engineered; the critical open question is whether such dynamics can emerge intrinsically within ANN architectures themselves. To address this, we implemented adaptation-like mechanisms directly within object recognition networks (Figure 5A). Specifically, we parameterized unit activations with exponential decay functions, using time constants derived from IT neuronal recordings during motion adaptation (Figure 5B). This allowed us to simulate repetition suppression across model units in a manner constrained by biological data. We first verified that these intrinsic suppression mechanisms reduced unit activity over simulated adaptation time (Supplementary Figure 2). Increasing adaptation strength by increasing the duration of the adapter progressively degraded the correlation between predicted object positions before and after motion adaptation (rightward: Mean $r_x$ = 0.69, SD = 0.13, leftward: Mean $r_x$ = 0.69, SD = 0.13) to a level similar to that seen in neural predictions (rightward: t(4) = -0.17, p = 0.87, single-sample t-test, leftward: t(4) = -0.54, p = 0.62, single-sample t-test; Figure 5C). However, despite reproducing IT-like suppression dynamics, these models did not generate systematic direction-opponent biases in position decodes. After simulated adaptation, predicted deltas in x-position ($\Delta P_x = P_{x,\ adapt\ left} - P_{x,\ adapt\ right}$) were near zero across AlexNet (Mean $\Delta P_x$ = -0.00022°, SD = 0.0011°), VGG-16 (Mean $\Delta P_x$ = 0.00058°, SD = 0.00077°), ResNet-18 (Mean $\Delta P_x$ = 0.0018°, SD = 0.00080°), ViT-L32 (Mean $\Delta P_x$ = 0.00032°, SD = 0.00045°) and SimCLR ResNet-50 (Mean $\Delta P_x$ = 0.0012°, SD = 0.00075°) and significantly smaller than the biases observed in humans (Mean $\Delta P_x$ = 0.33°, SD = 0.23°) and macaque IT (Mean $\Delta P_x$ = 0.84°, SD = 0.84°; models vs humans: AlexNet: t(39) = -8.81, p < 0.001, related samples t-test, VGG-16: t(39) = -8.79, p < 0.001, related samples t-test, ResNet-18: t(39) = -8.75, p < 0.001, related samples t-test, ViT-L32: t(39) = -8.79, p < 0.001, related samples t-test, SimCLR ResNet-50: t(39) = -8.77, p < 0.001, related samples t-test; models vs macaque IT: AlexNet: t(39) = -6.25, p < 0.001, related samples t-test, VGG-16: t(39) = -6.25, p < 0.001, related samples t-test, ResNet-18: t(39) = -6.23, p < 0.001, related samples t-test, ViT-L32: t(39) = -6.24, p < 0.001, related samples t-test, SimCLR ResNet-50: t(39) = -6.24, p < 0.001, related samples t-test; Figure 5D). These results show that while intrinsic decay functions capture the suppressive *signature* of adaptation, they are insufficient to account for the representational shifts that produce perceptual biases in IT. In other words, suppression alone cannot explain why adaptation induces direction-selective displacement in neural codes for position.

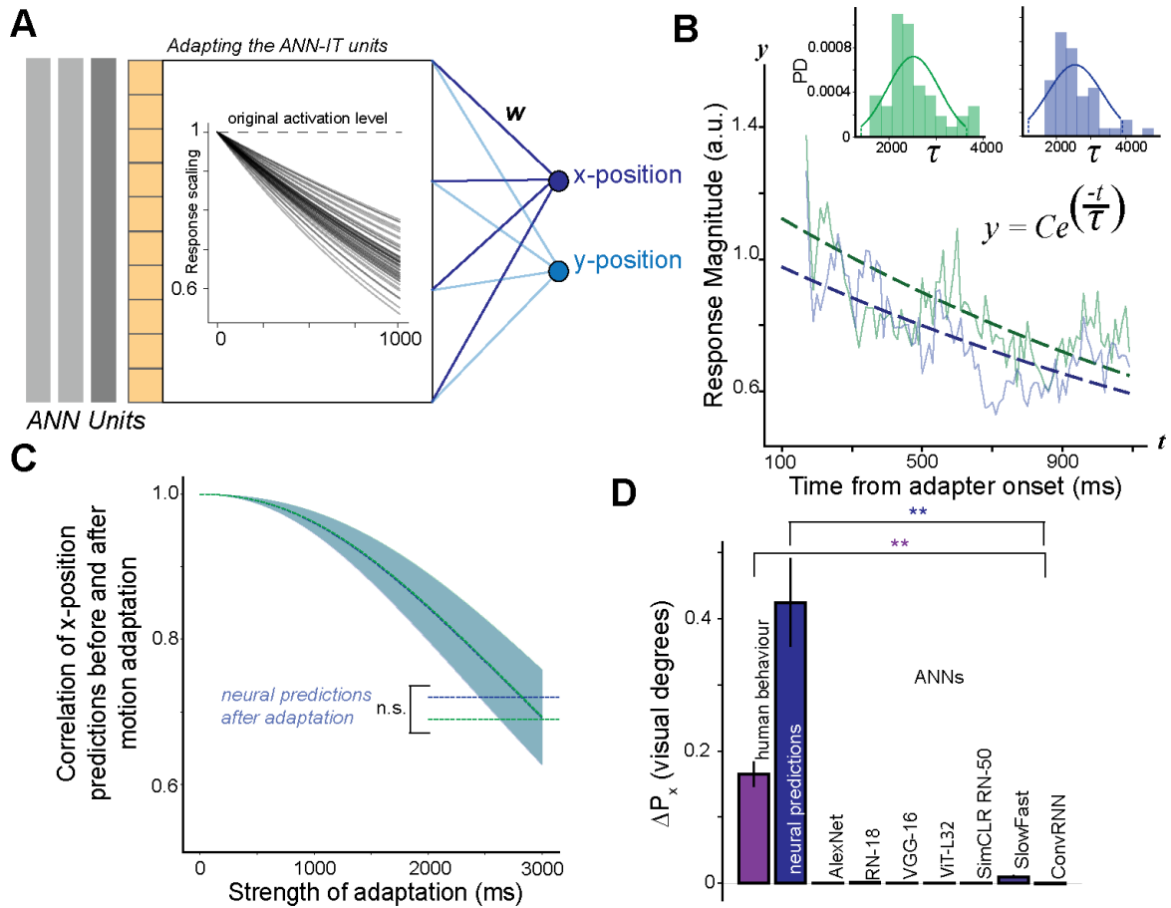

**Figure 5. Implementing intrinsic suppression dynamics in ANN units fails to recapitulate IT-like position biases. (A)** Schematic of simulated adaptation: unit activations in object recognition networks were parameterized with exponential decay functions fit to IT neuronal responses. Inset shows example simulated decay trajectories across ANN units. **(B)** An example neuronal unit ($\tau_{adapt\ right}$ = 1798.36, $\tau_{adapt\ left}$ = 1993.98) demonstrating that IT recordings during adapter presentation exhibited repetition suppression well fit by exponential functions ($\tau$ distributions shown in insets). **(C)** Correlation between predicted x-positions before and after motion adaptation (green: right, blue: left) as a function of adaptation strength. Increasing adaptation progressively reduced information content in ANN features, mimicking the suppressive effect observed in IT. Shaded region denotes error across models (standard error of the mean). **(D)** Comparison of position biases ($\Delta P_x = P_{x,\ adapt\ left} - P_{x,\ adapt\ right}$) across humans, macaque IT ("neural predictions"), and ANN models. Unlike IT and behavior, ANN models with intrinsic suppression and temporal convolution mechanisms (SlowFast and ConvRNN) failed to produce systematic direction-opponent shifts in position estimates. Error bars denote standard error of the mean across images; all differences between human estimates and model predictions, and neural and model predictions are significant ** $p < 0.001$.

## Temporal convolution in action recognition models does not reproduce adaptation-induced biases

Although embedding suppression dynamics into feedforward models reduced information content, it failed to generate perceptual biases (Figure 5). One possibility is that the missing ingredient is not suppression per se, but intrinsic temporal processing. Unlike feedforward object recognition networks, action recognition models such as SlowFast[54] and ConvRNN[53] are trained on video data and implement temporal convolutions and recurrent feedback, which may naturally produce history-dependent representations. We therefore tested whether these models generate adaptation-induced position shifts when exposed to prolonged motion stimuli. For the SlowFast architecture, we presented adapter movies of drifting gratings followed by stationary object images, and extracted features from IT-aligned layers in the slow pathway (Supplementary Figure 3A). Position decoders trained on pre-adaptation features generalized well to post-adaptation features (right-adapted: $r_x$ = 0.95, left-adapted: $r_x$ = 0.95). However, scatterplots of predicted positions revealed no systematic displacements for either x- or y-dimensions (Supplementary Figure 3B), and change in predictions between rightward and leftward adaptation ($\Delta P_x = P_{x, adapt\ left} - P_{x, adapt\ right}$) was significantly smaller than that observed in human estimates (Mean $\Delta Px$ = 0.33°, SD = 0.23°) and decodes from macaque IT (Mean $\Delta P_x$ = 0.84°, SD = 0.84°; SlowFast Mean $\Delta P_x$ = 0.019°, SD = 0.024°; SlowFast vs humans $t(39)$ = -8.36, $p < 0.001$, related samples t-test; SlowFast vs macaques $t(39)$ = -6.07, $p < 0.001$, repeated samples t-test; Figure 5D), indicating that temporal convolutions in SlowFast did not mimic the perceptually aligned shifts observed in IT. We next tested the ConvRNN model, which incorporates local recurrence and long-range feedback (Supplementary Figure 3C). Stimuli were structured to match the temporal regime of our macaque recordings (3 s adaptation, 200 ms gray interval, and 100 ms test presentation). Despite these biologically inspired dynamics, ConvRNN predictions again showed no direction-opponent bias. As with SlowFast, scatterplots aligned closely with the unity line for both x- and y-positions (Supplementary Figure 3D), confirming that adaptation failed to alter decoded position (Mean $\Delta P_x$ = -0.0016°, SD = 0.0050°) like in human estimates and predictions from macaque IT (ConvRNN vs humans $t(39)$ = -8.81, $p < 0.001$, ConvRNN vs macaques $t(39)$ = -6.25, $p < 0.001$; Figure 5D). Together, these findings indicate that temporal convolutions and recurrence, as currently implemented in state-of-the-art action recognition networks, are insufficient to capture the perceptually relevant adaptation dynamics expressed in the IT cortex. While these models exhibit history-dependent computations, they lack the mechanisms required to induce systematic positional aftereffects from identical pixel inputs.

# Discussion

In this study, we set out to test whether object position information in macaque inferior temporal (IT) cortex is perceptually aligned or merely a by-product of feedforward retinotopy. By leveraging the motion aftereffect (MAE), a classic paradigm that dissociates perceived from pixel-based object position, we created conditions where ground-truth and perceptual estimates diverge. Our results provide several key insights. First, IT responses contain decodable object position information that aligns with prior work[23]. Second, motion adaptation reshapes IT population geometry and biases position decodes in the direction opposite to adapter motion, qualitatively mirroring human perceptual reports. Finally, while activations from ANN models of the ventral stream can be "neuralized" to reproduce these biases by imposing empirically derived transformations, neither feedforward models with simulated suppression nor state-of-the-art temporal networks generated such effects intrinsically. Together, these results advance our understanding of IT as a perceptually aligned substrate for spatial coding, and highlight specific gaps between biological and artificial vision systems. Importantly, the position biases we observed cannot be explained by changes in retinal input. In our paradigm, the stationary test image was identical across adaptation conditions, meaning that any shift in decoded position must arise from changes in the neural representation itself rather than from differences in the stimulus. The fact that linear decoders trained on pre-adaptation responses systematically shifted their position estimates after adaptation therefore indicates that motion history alters the geometry of the IT population code. In other words, adaptation modifies how object position is represented rather than simply scaling neural responses.

## IT as a locus of perceptually aligned object position coding

Our findings strengthen the case that the IT cortex within the ventral stream encodes not only object identity but also position information in a manner directly relevant to perception. Classic two-stream models emphasize IT as the terminus of ventral "what" processing[2], yet growing evidence has shown that IT carries information about category-orthogonal properties such as size, pose, and position[23]. The present results strengthen that inference by providing a critical test in conditions where perceived and retinal positions diverge. The fact that IT position decodes shifted systematically with perceptual biases rules out the hypothesis that these signals simply reflect inherited retinotopy from earlier areas[33]. Crucially, this conclusion follows from the dissociation created by the motion aftereffect. Because the pixel-based location of the object remained constant across conditions, a purely feedforward retinotopic representation would predict identical position estimates before and after adaptation. Instead, the decoded positions shifted systematically in the direction opposite to the adapter motion, mirroring the perceptual biases observed in human observers. This alignment between neural readouts and behavior suggests that IT representations are already partially transformed into perceptually relevant spatial coordinates.

This conclusion dovetails with emerging evidence that IT participates in more dynamic and integrative computations than previously assumed. IT neurons have been shown to respond to motion trajectories[47], moving bodies[48], and object motion even in the absence of explicit form cues[49]. The present results push this view further, showing that IT responses are not only modulated by motion but also remapped in ways that underlie perceptual position shifts. This highlights IT as a site where ventral stream representations flexibly incorporate context, potentially through interactions with motion-sensitive regions such as MT and V4.

Our findings also reinforce that, while IT represents perceptually relevant information about object position, it is not necessarily the locus from which this information is directly read out to produce behaviour. While biases in predicted position emerged in both human estimates and decodes from macaque IT, in IT these biases were accompanied by a drop in correlation between predictions before and after motion adaptation that was not seen in human behaviour. This result suggests that other areas, downstream of IT, introduce novel transformations on object position representations that further influence behaviour in this illusion paradigm[58,59].

**Adaptation as a probe of circuit mechanisms**

Adaptation paradigms provide a powerful means of uncovering circuit-level mechanisms of sensory coding. In MT, the neural basis of MAE has been linked to direction-selective neurons undergoing repetition suppression[38,39]. Our IT recordings show analogous suppression during motion adaptation. This suggests that adaptation is not a uniform modulation but a selective re-weighting of signals within the IT population, shifting the geometry of the representational space.

At the circuit level, several mechanisms could contribute to this effect. One possibility is feedforward inheritance from motion-tuned inputs via V4, which receives strong projections from MT[60]. Suppressed inputs could alter IT receptive field balance, displacing apparent object centers. Another possibility is local recurrent dynamics within IT, which have been shown to enhance selectivity and sharpen discriminability during adaptation[50,51]. Finally, feedback from higher-order areas (e.g., prefrontal or parietal cortex) could reshape IT activity based on perceptual expectations during adaptation[58]. Teasing apart these possibilities will require targeted experiments, such as reversible inactivation of MT or V4 during adaptation, laminar recordings in IT to track feedforward vs. feedback influences, and perturbation of recurrent circuits using pharmacological or optogenetic tools.

**Why ANNs fail to capture perceptually aligned adaptation**

Our modeling experiments provide a complementary perspective. Standard feedforward ANNs, by design, predict stable position outputs for identical pixel inputs, regardless of perceptual context. This insensitivity underscores their lack of intrinsic dynamics. Notably, this failure does not arise because ANN features lack position information. In our baseline analyses, object position could be decoded from several architectures with accuracy comparable to or exceeding that of IT. The limitation therefore lies not in representational capacity but in the absence of dynamics that reshape the representational geometry based on sensory history. In biological populations, adaptation selectively reweights responses across neurons, whereas most current ANN architectures treat each input independently without incorporating comparable history-dependent gain control. Neuralization simulations showed that empirical IT-derived transformations on the ANN features are sufficient to impose perceptually aligned biases on ANN features, indicating that existing feature spaces are *capable* of representing such shifts. What is missing are mechanisms that naturally generate these transformations.

We tested two classes of candidate mechanisms. First, embedding exponential suppression in unit activations reproduced the firing-rate decay observed in IT but failed to produce systematic positional biases. This suggests that suppression alone is not sufficient: what matters is not just that units adapt, but how adaptation is structured across the population to alter representational geometry in a direction-specific manner. Second, we examined dynamic video models such as SlowFast[54] and ConvRNN[53], which implement temporal convolution and recurrent feedback. Despite their temporal richness, these models likewise failed to generate position biases, indicating that current architectures and training objectives do not produce the required context-dependent reweighting.

These negative results are instructive. They suggest that the key missing ingredient may be interaction between motion- and form-selective channels, coupled with history-dependent gain control. Unlike ANNs trained on static object classification or generic action recognition, the primate ventral stream evolved under constraints of perceptual stability in dynamic environments. Capturing this in silico may require architectures that explicitly couple form and motion pathways, trained with objectives that demand perceptual alignment rather than mere label prediction.

**Future directions: neural circuits**

Our data point toward several promising directions for circuit-level investigation. First, simultaneous recordings in IT and MT/V4 during adaptation would clarify the degree of feedforward inheritance versus local computation. Second, causal perturbations (e.g., chemogenetic silencing[61]) targeted at motion-sensitive regions versus form-sensitive subpopulations within IT could test whether positional biases depend on specific local circuits. Finally, cross-area coherence analyses could reveal whether adaptation reorganizes communication between ventral and dorsal streams, consistent with growing evidence of their interaction in dynamic vision.

**Future directions: artificial neural networks**

On the modeling side, our findings highlight the need for next-generation ANNs that move beyond static feedforward architectures[62]. Several design principles emerge. First, models should couple form and motion pathways rather than treating them as independent tasks; incorporating MT-like modules that project into IT-like layers may better capture perceptual alignment. Second, suppression must be structured rather than uniform, suggesting a role for mechanisms analogous to divisive normalization, tuned adaptation, or context-sensitive gating to reweight population codes in a manner consistent with IT. Third, recurrence and feedback loops need to be harnessed in ways that demand perceptual consistency, potentially through frameworks that integrate top-down signals. Fourth, training objectives must shift from categorical accuracy alone to perceptually aligned tasks, such as position estimation under adaptation or disambiguation of ambiguous stimuli, which would push networks to internalize the relevant dynamics. Finally, neurophysiological constraints should be embedded directly into models: as demonstrated by the neuralization approach, biological data can guide transformations[63,64], and fitting ANN dynamics to neuronal adaptation trajectories could yield models that perform well while also aligning mechanistically with cortical computations.

**Broader implications**

The present findings bridge several longstanding divides. At the theoretical level, they challenge strict interpretations of the two-stream hypothesis by showing that IT position codes are perceptually relevant. At the mechanistic level, they link population-level representational shifts to perceptual biases. At the modeling level, they identify both the potential and the limitations of current ANN architectures for explaining dynamic perceptual phenomena. More broadly, these results speak to a core problem of perceptual stability in dynamic environments. The brain must reconcile rapidly changing sensory input with relatively stable percepts. Adaptation-induced biases, rather than reflecting failure, may be signatures of this reconciliation process: by dynamically reweighting representations, the visual system aligns neural codes with perceptual experience. Understanding how IT contributes to this process not only advances vision science but also provides a blueprint for building artificial systems with human-like perceptual robustness. Our study demonstrates that IT encodes object position in perceptually aligned coordinates, that adaptation reshapes IT population geometry in ways that mirror human perceptual biases, and that current ANN models fail to reproduce these dynamics intrinsically. These findings advance our understanding of IT as a hub for integrating form and spatial information and highlight the need for new computational frameworks that couple motion and form, implement history-dependent gain control, and optimize for perceptual alignment. Teasing apart the circuit-level mechanisms that support these computations in the primate brain, and embedding their principles into artificial networks, will be crucial steps toward closing the gap between biological and artificial vision.

# References

1. Treisman, A. (1996). The binding problem. Curr. Opin. Neurobiol. *6*, 171–178. https://doi.org/10.1016/S0959-4388(96)80070-5.

2. Goodale, M.A., and Milner, A.D. (1992). Separate visual pathways for perception and action. Trends Neurosci. *15*, 20–25. https://doi.org/10.1016/0166-2236(92)90344-8.

3. Baizer, J., Ungerleider, L., and Desimone, R. (1991). Organization of visual inputs to the inferior temporal and posterior parietal cortex in macaques. J. Neurosci. *11*, 168–190. https://doi.org/10.1523/JNEUROSCI.11-01-00168.1991.

4. Morel, A., and Bullier, J. (1990). Anatomical segregation of two cortical visual pathways in the macaque monkey. Vis. Neurosci. *4*, 555–578. https://doi.org/10.1017/S0952523800005769.

5. Faugier-Grimaud, S., Frenois, C., and Stein, D.G. (1978). Effects of posterior parietal lesions on visually guided behavior in monkeys. Neuropsychologia *16*, 151–168. https://doi.org/10.1016/0028-3932(78)90103-3.

6. Miller, E.K., Gochin, P.M., and Gross, C.G. (1991). Habituation-like decrease in the responses of neurons in inferior temporal cortex of the macaque. Vis. Neurosci. *7*, 357–362. https://doi.org/10.1017/S0952523800004843.

7. Perenin, M.-T., and Vighetto, A. (1988). Optic Ataxia: A specific disruption in visuomotor mechanisms: I. Different Aapects Of The Deficit In Reaching For Objects. Brain *111*, 643–674. https://doi.org/10.1093/brain/111.3.643.

8. Weiskrantz, L., and Saunders, R.C. (1984). Impairments of visual object transforms in monkeys. Brain *107*, 1033–1072. https://doi.org/10.1093/brain/107.4.1033.

9. Bao, P., She, L., McGill, M., and Tsao, D.Y. (2020). A map of object space in primate inferotemporal cortex. Nature *583*, 103–108. https://doi.org/10.1038/s41586-020-2350-5.

10. DiCarlo, J.J., Zoccolan, D., and Rust, N.C. (2012). How Does the Brain Solve Visual Object Recognition? Neuron *73*, 415–434. https://doi.org/10.1016/j.neuron.2012.01.010.

11. Downing, P.E., Jiang, Y., Shuman, M., and Kanwisher, N. (2001). A Cortical Area Selective for Visual Processing of the Human Body. Science *293*, 2470–2473. https://doi.org/10.1126/science.1063414.

12. Epstein, R., Harris, A., Stanley, D., and Kanwisher, N. (1999). The Parahippocampal Place Area. Neuron *23*, 115–125. https://doi.org/10.1016/S0896-6273(00)80758-8.

13. Kanwisher, N., McDermott, J., and Chun, M.M. (1997). The Fusiform Face Area: A Module in Human Extrastriate Cortex Specialized for Face Perception. J. Neurosci. *17*, 4302–4311. https://doi.org/10.1523/JNEUROSCI.17-11-04302.1997.

14. Kornblith, S., Cheng, X., Ohayon, S., and Tsao, D.Y. (2013). A Network for Scene Processing in the Macaque Temporal Lobe. Neuron *79*, 766–781. https://doi.org/10.1016/j.neuron.2013.06.015.

15. Popivanov, I.D., Jastorff, J., Vanduffel, W., and Vogels, R. (2012). Stimulus representations in body-selective regions of the macaque cortex assessed with event-related fMRI. NeuroImage *63*, 723–741. https://doi.org/10.1016/j.neuroimage.2012.07.013.

16. Schwarzlose, R.F., Baker, C.I., and Kanwisher, N. (2005). Separate Face and Body Selectivity on the Fusiform Gyrus. J. Neurosci. *25*, 11055–11059. https://doi.org/10.1523/JNEUROSCI.2621-05.2005.

17. Tsao, D.Y., Freiwald, W.A., Tootell, R.B.H., and Livingstone, M.S. (2006). A Cortical Region Consisting Entirely of Face-Selective Cells. Science *311*, 670–674. https://doi.org/10.1126/science.1119983.

18. Andersen, R.A. (2022). The Neurobiological Basis of Spatial Cognition: Role of the Parietal Lobe. In Spatial Cognition (Psychology Press), pp. 57–80. https://doi.org/10.4324/9781315785462-4.

19. Bracci, S., Daniels, N., and Op De Beeck, H. (2017). Task Context Overrules Object- and Category-Related Representational Content in the Human Parietal Cortex. Cereb. Cortex, cercor;bhw419v1. https://doi.org/10.1093/cercor/bhw419.

20. Chao, L.L., and Martin, A. (2000). Representation of Manipulable Man-Made Objects in the Dorsal Stream. NeuroImage *12*, 478–484. https://doi.org/10.1006/nimg.2000.0635.

21. Aflalo, T., Kellis, S., Klaes, C., Lee, B., Shi, Y., Pejsa, K., Shanfield, K., Hayes-Jackson, S., Aisen, M., Heck, C., et al. (2015). Decoding motor imagery from the posterior parietal cortex of a tetraplegic human. Science *348*, 906–910. https://doi.org/10.1126/science.aaa5417.

22. Bonner, M.F., and Epstein, R.A. (2017). Coding of navigational affordances in the human visual system. Proc. Natl. Acad. Sci. *114*, 4793–4798. https://doi.org/10.1073/pnas.1618228114.

23. Hong, H., Yamins, D.L.K., Majaj, N.J., and DiCarlo, J.J. (2016). Explicit information for category-orthogonal object properties increases along the ventral stream. Nat. Neurosci. *19*, 613–622. https://doi.org/10.1038/nn.4247.

24. Peelen, M.V., and Caramazza, A. (2012). Conceptual Object Representations in Human Anterior Temporal Cortex. J. Neurosci. *32*, 15728–15736. https://doi.org/10.1523/JNEUROSCI.1953-12.2012.

25. Freud, E., Plaut, D.C., and Behrmann, M. (2016). 'What' Is Happening in the Dorsal Visual Pathway. Trends Cogn. Sci. *20*, 773–784. https://doi.org/10.1016/j.tics.2016.08.003.

26. Vaina, L.M., Solomon, J., Chowdhury, S., Sinha, P., and Belliveau, J.W. (2001). Functional neuroanatomy of biological motion perception in humans. Proc. Natl. Acad. Sci. *98*, 11656–11661. https://doi.org/10.1073/pnas.191374198.

27. Robert, S., Ungerleider, L.G., and Vaziri-Pashkam, M. (2023). Disentangling Object Category Representations Driven by Dynamic and Static Visual Input. J. Neurosci. *43*, 621–634. https://doi.org/10.1523/JNEUROSCI.0371-22.2022.

28. Tse, P.U. (2006). Neural correlates of transformational apparent motion. NeuroImage *31*, 766–773. https://doi.org/10.1016/j.neuroimage.2005.12.029.

29. Mahon, B.Z., and Almeida, J. (2024). Reciprocal interactions among parietal and occipito-temporal representations support everyday object-directed actions. Neuropsychologia *198*, 108841. https://doi.org/10.1016/j.neuropsychologia.2024.108841.

30. Milner, A.D. (2017). How do the two visual streams interact with each other? Exp. Brain Res. *235*, 1297–1308. https://doi.org/10.1007/s00221-017-4917-4.

31. Majaj, N.J., Hong, H., Solomon, E.A., and DiCarlo, J.J. (2015). Simple Learned Weighted Sums of Inferior Temporal Neuronal Firing Rates Accurately Predict Human Core Object Recognition Performance. J. Neurosci. *35*, 13402–13418. https://doi.org/10.1523/JNEUROSCI.5181-14.2015.

32. Kar, K., and DiCarlo, J.J. (2024). The Quest for an Integrated Set of Neural Mechanisms Underlying Object Recognition in Primates. Annu. Rev. Vis. Sci. *10*, 91–121. https://doi.org/10.1146/annurev-vision-112823-030616.

33. Conway, B.R. (2018). The Organization and Operation of Inferior Temporal Cortex. Annu. Rev. Vis. Sci. *4*, 381–402. https://doi.org/10.1146/annurev-vision-091517-034202.

34. Yamins, D.L.K., Hong, H., Cadieu, C.F., Solomon, E.A., Seibert, D., and DiCarlo, J.J. (2014). Performance-optimized hierarchical models predict neural responses in higher visual cortex. Proc. Natl. Acad. Sci. *111*, 8619–8624. https://doi.org/10.1073/pnas.1403112111.

35. Rajalingham, R., Issa, E.B., Bashivan, P., Kar, K., Schmidt, K., and DiCarlo, J.J. (2018). Large-Scale, High-Resolution Comparison of the Core Visual Object Recognition Behavior of Humans, Monkeys, and State-of-the-Art Deep Artificial Neural Networks. J. Neurosci. *38*, 7255–7269. https://doi.org/10.1523/JNEUROSCI.0388-18.2018.

36. Anstis, S., Verstraten, F.A.J., and Mather, G. (1998). The motion aftereffect. Trends Cogn. Sci. *2*, 111–117. https://doi.org/10.1016/S1364-6613(98)01142-5.

37. Born, R.T., and Bradley, D.C. (2005). Structure and Function of Visual Area MT. Annu. Rev. Neurosci. *28*, 157–189. https://doi.org/10.1146/annurev.neuro.26.041002.131052.

38. Petersen, S.E., Baker, J.F., and Allman, J.M. (1985). Direction-specific adaptation in area MT of the owl monkey. Brain Res. *346*, 146–150. https://doi.org/10.1016/0006-8993(85)91105-9.

39. Huk, A.C., Ress, D., and Heeger, D.J. (2001). Neuronal Basis of the Motion Aftereffect Reconsidered. Neuron *32*, 161–172. https://doi.org/10.1016/S0896-6273(01)00452-4.

40. Kar, K., and Krekelberg, B. (2016). Testing the assumptions underlying fMRI adaptation using intracortical recordings in area MT. Cortex *80*, 21–34. https://doi.org/10.1016/j.cortex.2015.12.011.

41. Ferrera, V., Rudolph, K., and Maunsell, J. (1994). Responses of neurons in the parietal and temporal visual pathways during a motion task. J. Neurosci. *14*, 6171–6186. https://doi.org/10.1523/JNEUROSCI.14-10-06171.1994.

42. Li, P., Zhu, S., Chen, M., Han, C., Xu, H., Hu, J., Fang, Y., and Lu, H.D. (2013). A Motion Direction Preference Map in Monkey V4. Neuron *78*, 376–388. https://doi.org/10.1016/j.neuron.2013.02.024.

43. Mysore, S.G., Vogels, R., Raiguel, S.E., and Orban, G.A. (2006). Processing of Kinetic Boundaries in Macaque V4. J. Neurophysiol. *95*, 1864–1880. https://doi.org/10.1152/jn.00627.2005.

44. Tolias, A.S., Keliris, G.A., Smirnakis, S.M., and Logothetis, N.K. (2005). Neurons in macaque area V4 acquire directional tuning after adaptation to motion stimuli. Nat. Neurosci. *8*, 591–593. https://doi.org/10.1038/nn1446.

45. Sary, G., Vogels, R., Kovacs, G., and Orban, G.A. (1995). Responses of monkey inferior temporal neurons to luminance-, motion-, and texture-defined gratings. J. Neurophysiol. *73*, 1341–1354. https://doi.org/10.1152/jn.1995.73.4.1341.

46. Vanduffel, W., Fize, D., Mandeville, J.B., Nelissen, K., Van Hecke, P., Rosen, B.R., Tootell, R.B.H., and Orban, G.A. (2001). Visual Motion Processing Investigated Using Contrast Agent-Enhanced fMRI in Awake Behaving Monkeys. Neuron *32*, 565–577. https://doi.org/10.1016/S0896-6273(01)00502-5.

47. Burk, D.C., and Sheinberg, D.L. (2022). Neurons in inferior temporal cortex are sensitive to motion trajectory during degraded object recognition. Cereb. Cortex Commun. *3*, tgac034. https://doi.org/10.1093/texcom/tgac034.

48. Raman, R., Bognár, A., Nejad, G.G., Taubert, N., Giese, M., and Vogels, R. (2023). Bodies in motion: Unraveling the distinct roles of motion and shape in dynamic body responses in the temporal cortex. Cell Rep. *42*, 113438. https://doi.org/10.1016/j.celrep.2023.113438.

49. Ramezanpour, H., Ilic, F., Wildes, R.P., and Kar, K. (2024). Object motion representation in the macaque ventral stream – a gateway to understanding the brain's intuitive physics engine. Preprint at Neuroscience, https://doi.org/10.1101/2024.02.23.581841 https://doi.org/10.1101/2024.02.23.581841.

50. Vogels, R. (2016). Sources of adaptation of inferior temporal cortical responses. Cortex *80*, 185–195. https://doi.org/10.1016/j.cortex.2015.08.024.

51. Kaliukhovich, D.A., De Baene, W., and Vogels, R. (2013). Effect of Adaptation on Object Representation Accuracy in Macaque Inferior Temporal Cortex. J. Cogn. Neurosci. *25*, 777–789. https://doi.org/10.1162/jocn_a_00355.

52. Schrimpf, M., Kubilius, J., Hong, H., Majaj, N.J., Rajalingham, R., Issa, E.B., Kar, K., Bashivan, P., Prescott-Roy, J., Geiger, F., et al. (2018). Brain-Score: Which Artificial Neural Network for Object Recognition is most Brain-Like? Preprint at Neuroscience, https://doi.org/10.1101/407007 https://doi.org/10.1101/407007.

53. Nayebi, A., Sagastuy-Brena, J., Bear, D.M., Kar, K., Kubilius, J., Ganguli, S., Sussillo, D., DiCarlo, J.J., and Yamins, D.L.K. (2022). Recurrent Connections in the Primate Ventral Visual Stream Mediate a Trade-Off Between Task Performance and Network Size During Core Object Recognition. Neural Comput. *34*, 1652–1675. https://doi.org/10.1162/neco_a_01506.

54. Feichtenhofer, C., Fan, H., Malik, J., and He, K. (2019). SlowFast Networks for Video Recognition. Preprint at arXiv, https://doi.org/10.48550/arXiv.1812.03982 https://doi.org/10.48550/arXiv.1812.03982.

55. Vinken, K., Boix, X., and Kreiman, G. (2020). Incorporating intrinsic suppression in deep neural networks captures dynamics of adaptation in neurophysiology and perception. Sci. Adv. *6*, eabd4205. https://doi.org/10.1126/sciadv.abd4205.

56. Kornblith, S., Norouzi, M., Lee, H., and Hinton, G. (2019). Similarity of Neural Network Representations Revisited. Preprint at arXiv, https://doi.org/10.48550/ARXIV.1905.00414 https://doi.org/10.48550/ARXIV.1905.00414.

57. Muzellec, S., and Kar, K. (2025). Reverse Predictivity: Going Beyond One-Way Mapping to Compare Artificial Neural Network Models and Brains. Preprint at Neuroscience, https://doi.org/10.1101/2025.08.08.669382 https://doi.org/10.1101/2025.08.08.669382.

58. Kar, K., and DiCarlo, J.J. (2021). Fast Recurrent Processing via Ventrolateral Prefrontal Cortex Is Needed by the Primate Ventral Stream for Robust Core Visual Object Recognition. Neuron *109*, 164-176.e5. https://doi.org/10.1016/j.neuron.2020.09.035.

59. Freedman, D.J., Riesenhuber, M., Poggio, T., and Miller, E.K. (2001). Categorical Representation of Visual Stimuli in the Primate Prefrontal Cortex. Science *291*, 312–316. https://doi.org/10.1126/science.291.5502.312.

60. Felleman, D.J., and Van Essen, D.C. (1991). Distributed Hierarchical Processing in the Primate Cerebral Cortex. Cereb. Cortex *1*, 1–47. https://doi.org/10.1093/cercor/1.1.1.

61. Kar, K. (2023). Probing the role of bypass connections in core object recognition by chemogenetic suppression of macaque V4 neurons. J. Vis. *23*, 5736. https://doi.org/10.1167/jov.23.9.5736.

62. Dunnhofer, M., Micheloni, C., and Kar, K. (2026). Better, But Not Sufficient: Testing Video ANNs Against Macaque IT Dynamics. Preprint at arXiv, https://doi.org/10.48550/arXiv.2601.03392 https://doi.org/10.48550/arXiv.2601.03392.

63. Dapello, J., Kar, K., Schrimpf, M., Geary, R.B., Ferguson, M., Cox, D.D., and DiCarlo, J.J. (2022). Aligning Model and Macaque Inferior Temporal Cortex Representations Improves Model-to-Human Behavioral Alignment and Adversarial Robustness. In The Eleventh International Conference on Learning Representations.

64. Chen, S., Cheng, Y.-A., Kar, K., Rodriguez, I., Serre, T., and Watanabe, T. (2024). RTify: Aligning Deep Neural Networks with Human Behavioral Decisions. In Advances in Neural Information Processing Systems 37 (Neural Information Processing Systems Foundation, Inc. (NeurIPS)), pp. 130485–130510. https://doi.org/10.52202/079017-4147.

65. He, K., Zhang, X., Ren, S., and Sun, J. (2016). Deep Residual Learning for Image Recognition. In 2016 IEEE Conference on Computer Vision and Pattern Recognition (CVPR) (IEEE), pp. 770–778. https://doi.org/10.1109/CVPR.2016.90.

66. Russakovsky, O., Deng, J., Su, H., Krause, J., Satheesh, S., Ma, S., Huang, Z., Karpathy, A., Khosla, A., Bernstein, M., et al. (2015). ImageNet Large Scale Visual Recognition Challenge. Int. J. Comput. Vis. *115*, 211–252. https://doi.org/10.1007/s11263-015-0816-y.

67. Chen, T., Kornblith, S., Norouzi, M., and Hinton, G. (2020). A Simple Framework for Contrastive Learning of Visual Representations. Preprint at arXiv, https://doi.org/10.48550/ARXIV.2002.05709 https://doi.org/10.48550/ARXIV.2002.05709.

68. Caron, M., Bojanowski, P., Joulin, A., and Douze, M. (2018). Deep Clustering for Unsupervised Learning of Visual Features. Preprint at arXiv, https://doi.org/10.48550/ARXIV.1807.05520 https://doi.org/10.48550/ARXIV.1807.05520.

69. Salman, H., Yang, G., Li, J., Zhang, P., Zhang, H., Razenshteyn, I., and Bubeck, S. (2019). Provably Robust Deep Learning via Adversarially Trained Smoothed Classifiers. Preprint at arXiv, https://doi.org/10.48550/ARXIV.1906.04584 https://doi.org/10.48550/ARXIV.1906.04584.

70. Krizhevsky, A., Sutskever, I., and Hinton, G.E. (2012). ImageNet Classification with Deep Convolutional Neural Networks. In Advances in Neural Information Processing Systems, F. Pereira, C. J. C. Burges, L. Bottou, and K. Q. Weinberger, eds. (Curran Associates, Inc.).

71. Simonyan, K., Vedaldi, A., and Zisserman, A. (2014). Deep Inside Convolutional Networks: Visualising Image Classification Models and Saliency Maps. Preprint at arXiv, https://doi.org/10.48550/arXiv.1312.6034 https://doi.org/10.48550/arXiv.1312.6034.

72. Dosovitskiy, A., Beyer, L., Kolesnikov, A., Weissenborn, D., Zhai, X., Unterthiner, T., Dehghani, M., Minderer, M., Heigold, G., Gelly, S., et al. (2021). An Image is Worth 16x16 Words: Transformers for Image Recognition at Scale. Preprint at arXiv, https://doi.org/10.48550/ARXIV.2010.11929 https://doi.org/10.48550/ARXIV.2010.11929.

73. Kay, W., Carreira, J., Simonyan, K., Zhang, B., Hillier, C., Vijayanarasimhan, S., Viola, F., Green, T., Back, T., Natsev, P., et al. (2017). The Kinetics Human Action Video Dataset. Preprint at arXiv, https://doi.org/10.48550/ARXIV.1705.06950 https://doi.org/10.48550/ARXIV.1705.06950.

# Material and Methods

**Visual Stimuli**

HVM640

A sample of 640 images from the 5,760 imageset used in Hong et al. (2016)[23]. All images were achromatic, with a native resolution of 256 x 256 pixels covering 8° of foveal eccentricity. Each image contained one of 8 objects (bear, elephant, face, car, dog, apple, chair, plane) on a randomly selected naturalistic background. Object size varied from occluding 25% to 64% of the image on the longest axis. Objects were rotated between -90° and 90° of in-plane and out-of-plane rotation. Along the x-axis, the mean object position was at the center of the image, with a standard deviation of 0.28 scaled pixels (sps). The most leftward object had a position of 0.60 sps from the center of the image and the most rightward object had a position of -0.60 sps from the center of the image. Along the y-axis, the mean object position was 0.05 sps from the center of the image, with a standard deviation of 0.55 sps. The highest object had a position of 1.20 sps from the center of the image and the lowest object had a position of -1.20 sps from the center of the image.

MAE40

An image set (subset of HVM640) containing 40, achromatic images, with a native resolution of 256 x 256 pixels covering 8° of foveal eccentricity. Each image contained one of 8 objects (bear, elephant, face, car, dog, apple, chair, plane), on a randomly selected realistic background. Object size varied from occluding 25% to 64% of the image on the longest axis. Objects were rotated between -90° and 90° of in-plane and out-of-plane rotation. Along the x-axis, the mean object position was 125.025 pixels from the left-hand side of the image with a standard deviation of 20.76 pixels. The most leftward object had a position of 76 pixels from the left-hand side of the image and the most rightward object had a position of 184 pixels from the left-hand side of the image. Along the y-axis, the mean object position was 114.9 pixels from the top of the image, with a standard deviation of 38.88 pixels. The highest object had a position of 30 pixels from the top of the image and the lowest object had a position of 208 pixels from the top of the image.

Before being used for as ground truth labels for training object position regressors, values were rescaled to the dimensions on which human participants made behavioural estimates by multiplying by a factor of 256/p, where p refers to the size of the image during the behavioural experiment for each participant, and converted to visual degrees (°) using

$$d = \left(p - \frac{s}{2}\right)\left(\frac{16}{s}\right)$$

where $d$ is the position in visual degrees, $p$ is the position in pixels, and $s$ is the size of the image in pixels.

**Subjects**

## Human Participants

A total of 79 human subjects completed online experiments to test the effects of motion adaptation on object position estimation. Different participants completed each of the three conditions. 35 human participants completed the position estimation task following no adapter presentation. 22 human participants completed the position estimation task following adaptation to rightward motion. 22 human participants completed the position estimation task following adaptation to leftward motion. All participants were recruited from Amazon's Mechanical Turk pool of participants. Participants were selected to be between the ages of 18 and 45 years old, with no known un-corrected visual impairments, no seizures or known epileptic episodes, and no diagnosed neurological disorders.

## Non human primates

We have two adult male rhesus macaques (Macaca mulatta) as research subjects in our experiments. All data were collected, and animal procedures were performed, in accordance with the NIH guidelines, the Massachusetts Institute of Technology Committee on Animal Care, and the guidelines of the Canadian Council on Animal Care on the use of laboratory animals, and were also approved by the York University Animal Care Committee.

## Deep Artificial Neural Networks

We investigated the effects of motion adaptation on object position estimation in a set of n=9 deep artificial neural networks, all trained by other labs for engineering purposes. For each model, we collected unit responses from the layer that best aligns with brain data from the macaque IT cortex as determined during BrainScore region layer commitment[52]. Table 1 shows an overview of the models, and extensive documentation on the model architectures and training can be found in the original papers accompanying the models.

| Model Name | Architecture | IT-Like Layer | Objective | Training Regime | Training Dataset |
|---|---|---|---|---|---|
| ResNet-18[65] | Convolutional with skip connections, 18 layers | layer4.1.conv2 | Object recognition | Supervised | ImageNet [66] |
| SimCLR ResNet-50[67] | Convolutional with skip connections, 50 layers | layer3.2.bn1 | Object recognition | Self-supervised using Contrastive learning framework | ImageNet |
| ResNet-50 SSL[68] | Convolutional with skip connections, 50 layers | layer4.0 | Object recognition | Self-supervised using DeepCluster framework | ImageNet |
| Robust ResNet-50 (eps 3)[69] | Convolutional with skip connections, 50 layers | layer4.0 | Object recognition | Supervised, adversarially robust | ImageNet |
| AlexNet[70] | Convolutional, 9 layers | features.12 | Object recognition | Supervised | ImageNet |
| VGG-16 [71] | Convolutional, 16 layers | features.30 | Object recognition | Supervised | ImageNet |
| ViT-L32 [72] | Transformer | encoder.layers.encoder_layer_8.mlp | Object recognition | Supervised | ImageNet |
| SlowFast [54] | 3D convolutional with two streams, each stream is 50 layers with skip connections | blocks.4.multipathway_blocks.0.res_blocks.2.activation | Action recognition | Supervised | Kinetics-400[73] |
| ConvRNN [53] | Convolutional with local recurrent reciprocal gated circuits and long-range feedback, 10 layers | conv.10 | Object recognition | Supervised | ImageNet |

Table 1. ANN models used in the study.

## Models with Temporal Response Dynamics

As motion adaptation is widely recognized to rely on changes in neural responses over time [36], we included two models that were designed to implement unit response changes over time.

### SlowFast

SlowFast Network[54] for video action recognition, pre-trained on the Kinetics 400 Dataset, was downloaded from PyTorch Video. The model architecture is a temporally strided 3D ResNet-50, with parallel Slow and Fast streams with 5 ResStages each, fused to each other with lateral connections. When a video stimulus is presented to the model, both Slow and Fast streams sample video frames, and perform convolutions across 8 sampled time-points in the Slow stream and 32 sampled time-points in the Fast stream. Both streams also perform spatial convolutions, pixel-wise across 256 x 256 pixels of each sampled frame, with the spatial convolution channel capacity of the Fast stream being reduced to ¼ of the channels of the Slow stream. As such, the computations of the Slow stream allow it to extract semantic information from the videos, while the Fast stream monitors dynamic changes across frames. Responses from the two streams are pooled in the global average pool layer and a decoder outputs a label that classifies the action performed in the video.

The functional segregation of the SlowFast Network's streams, which loosely maps to the primate ventral and dorsal visual streams, as well as its strong performance as an action recognition model, makes it our selection for an initial test of the effects of motion adaptation on object position representations in ANN models of the visual system. As our aim was to probe object position information in the ventral stream specifically, we recorded unit responses from the first convolutional layer in the 5th ResStage in the Slow stream (Table 1) throughout adapter and stationary test image presentation.

### ConvRNN

The ConvRNN (convolutional recurrent neural network[53]) for object recognition, pre-trained on ImageNet, was downloaded from the original GitHub repository (https://github.com/neuroailab/convrnns) and adapted to process our dynamic stimuli instead of sequences composed of a single repeated image. ConvRNN is designed to computationally mimic the neural mechanisms of the primate visual ventral stream for core object recognition. The original model combines layer-local recurrent circuits within a feedforward convolutional neural network, with the option to include long-range feedback connections between layers. It is "biologically unrolled," meaning that signal propagation along each connection corresponds to one time step (~10 ms), mimicking conduction delays between cortical layers in the visual system. Specifically, we ran the rgc_intermediate configuration, as it is both highly performant in the object recognition task and closely matches the temporal dynamics of primate core object recognition data. We recorded activations from the conv10 layer (as specified in Table 1), which the authors recommend as the most analogous to anterior IT (aIT).

**Behavioral testing**

## Probing the reliability of object position estimates

After fixating on a white cross for 500 ms, a test image from MAE40 was presented for 100 ms. Participants reported the position of the object in the image by mouse-click on a blank screen following the image presentation. The x- and y-positions (pixels from the left and top of the image) of the mouse-click were recorded for each trial. There was an inter-trial interval of 1000 ms. The order of image presentation was randomized across participants.

## Probing the effect of motion adaptation on object position estimates

Participants fixated on a white cross while an adapter movie played, with gratings drifting either right or left, for 30 s. After this initial adaptation period, each trial consisted of 3000 ms of top-up adaptation to rightward or leftward motion, followed by an image from MAE40 presented for 100 ms, and a subsequent blank screen where participants reported the position of the object in the image by mouse-click. Each participant completed all trials preceded either by a right or left motion adapter. There was an inter-trial interval of 1000 ms. The order of MAE40 image presentation was randomized across participants.

**Primate passive fixation task**

## Neural Recordings

We surgically implanted each monkey with a head post under aseptic conditions. We recorded neural activity using two or three micro-electrode arrays (Utah arrays; Blackrock Microsystems) implanted in the IT cortex. A total of 96 electrodes were connected per array (grid arrangement, 400 um spacing, 4mm x 4mm span of each array). Array placement was guided by the sulcus pattern, which was visible during the surgery. The electrodes were accessed through a percutaneous connector that allowed simultaneous recording from all 96 electrodes from each array. All data were collected, and animal procedures were performed, in accordance with the NIH guidelines, the Massachusetts Institute of Technology Committee on Animal Care, and the guidelines of the Canadian Council on Animal Care on the use of laboratory animals and were also approved by the York University Animal Care Committee. During each daily recording session, band-pass filtered (0.1 Hz to 10 kHz) neural activity was recorded continuously at a sampling rate of 20 kHz using Intan Recording Controllers (Intan Technologies, LLC). The majority of the data presented here were based on multiunit activity.

## Eye Tracking

We monitored eye movements using video eye tracking (SR Research EyeLink 1000). Using operant conditioning and water reward, our 4 subjects were trained to fixate a central white square (0.2°) within a square fixation window that ranged from ±1°. At the start of each behavioral session, monkeys performed an eye-tracking calibration task by making a saccade to a range of spatial targets and maintaining fixation for 500 ms. Calibration was repeated if drift was noticed over the course of the session. Real-time eye-tracking was employed to ensure that eye jitter did not exceed ±2°, otherwise the trial was aborted, and data discarded. Stimulus display and reward control were managed using the MWorks Software (https://mworks.github.io).

### Probing object position representations in Macaque IT

A test image from HVM540 or MAE40 was presented for 100 ms while the animal maintained fixation. A water reward was delivered for appropriately maintained fixation.

### Probing the effect of motion adaptation on object position representations in Macaque IT

In each trial, an adapter movie was played, with gratings drifting either right or left, for 3000 ms. This was followed by a 300 ms blank screen and an image from MAE40 presented for 100 ms. All trials in one recording session were preceded either by a right or left motion adapter. There was an inter-trial interval of 1000 ms. A water reward was delivered for appropriately maintained fixation during the duration of each trial.

## Deep Artificial Neural Network In-Silico Experiments

### Testing object position information in static deep ANNs

To test for object position information in ANNs optimized for object recognition on ImageNet, we presented images from both HVM640 and MAE40 to a variety of static deep ANNs from Table 1 (all models except SlowFast and ConvRNN are considered static for the purposes of this study). Our choices of models reflect those that have set engineering benchmarks in the ImageNet challenge or represent significant architectural innovations in the computer vision community. To assess the effects of self-supervised training and adversarial robustness on position estimates, we include three variants of ResNet-50, two trained with different self-supervised learning frameworks and one with adversarial robustness (Table 1).

Before being passed to each model, images were pre-processed using a modified version of the standard ImageNet preprocessing pipeline. Specifically, each image was resized to 224 x 224 pixels, scaled to contain pixel values [0,1], and normalized such that image statistics were centered around [0.485, 0.456, 0.406] with standard deviations of [0.229, 0.224, 0.225] in the red, green, and blue color channels respectively.

### Neuralizing static deep ANNs

We analyzed neural population responses obtained under three adaptation conditions: no adaptation, rightward adaptation, and leftward adaptation. For each image, we also extracted deep neural network feature vectors from pretrained models and obtained corresponding spatial position labels. To construct neuralized feature spaces, we employed partial least squares regression (PLS) with eight latent components and 20-fold cross-validation at the image level. In each fold, network features from the training images were first linearly mapped into the neural response space of the no-adaptation condition. The predicted responses in this reference space were then transformed into the rightward- and leftward-adapted neural spaces using mappings derived from the empirical relationships between no-adaptation and adapted conditions. These transformed neural predictions were subsequently projected back into network feature space, yielding neuralized feature representations corresponding to rightward and leftward adaptation. A single linear decoder, trained only once to predict image position from network features in the no-adaptation condition, was then applied to both the baseline and neuralized features on held-out test images. This procedure ensured that all transformations (network-to-neural, neural-to-neural, and neural-to-network) and all position predictions were cross-validated, with training and test sets strictly separated. The full pipeline was repeated across multiple random seeds, and predictions were aggregated across repetitions to obtain stable estimates of position readout under each adaptation condition.

### Simulating motion adaptation in static deep ANNs

To examine whether the effects of motion adaptation on position estimates can be simulated in ANNs using a neural response decay parameter, we simulated motion adaptation in the model unit responses to MAE40 recorded during the first experiment. For each model unit, we selected a response decay parameter $\tau$ from the normal distribution of such parameters, estimated from macaque IT neuronal responses to left and right motion adapter stimuli. The response of the model unit at time $t$ following the onset of adapter presentation was then computed to be

$$y = Ce^{\left(\frac{-t}{\tau}\right)}$$

given the initial unit response $C$ when $t = 0$. The experiment was repeated 30 times to acquire a distribution of adapted model responses.

### Probing effects of motion adaptation on object position information in dynamic deep ANNs

#### SlowFast

To confirm the presence of explicitly available object position information in the Slow stream of the SlowFast Network, we presented the model with each of the 40 movies in MAE40. Subsequently, to test the effects of motion adaptation on object position representations, we presented the SlowFast Network with videos where a 3000 ms adapter, with gratings moving right or left, preceded one of the 40 movies from MAE40, presented for 100 ms. The model 'viewed' all 40 movies preceded by rightward adapters before viewing all 40 movies preceded by leftward adapters.

#### ConvRNN

When comparing model responses to neural data, Nayebi et al.[53] recommend presenting inputs in the same temporal structure as in the neural experiments: each time step in the model corresponds to ~10 ms of presentation. In their setup, this involves feeding the model a sequence of identical images for the duration of the visual presentation to the subject, followed by a sequence of blank (gray) images for the time steps when the stimulus is absent but neural recordings are still ongoing.

We modified the original code so that the input sequence consisted of 180 video frames containing a left or right motion adapter followed by 12 gray images, then 6 repetitions of the test image from MAE40. This was then followed by 50 gray images to remove the visual stimulus while continuing to record the model's responses. In temporal terms, this corresponds approximately to 3000 ms for the motion adaptation, 200 ms of gray, 100 ms for image presentation, and 500 ms of gray background, closely matching the presentation of stimuli to humans and non-human primates. Like SlowFast, ConvRNN viewed all 40 images from MAE40 in this manner, first preceded by rightward adapters then leftward adapters.

**Data Analyses**

## Behavioral metrics from humans

### Split-Half Reliability

Split-half reliability of position estimates by humans was computed across trials, where each trial is a completion of the entire behavioural experiment for a single condition by a single human participant. Trials for each condition were randomly split into two groups and averaged to create one mean estimate per image. Estimates of the two groups were correlated with each other (Pearson r). The correlation coefficient was corrected using the Spearman-Brown prediction formula. This process was bootstrapped (i.e., repeated) over 20 iterations and the average correlation coefficient across the 20 iterations was taken as the split-half reliability of the human position estimates for condition, with standard deviation computed to estimate the variance. This was done separately for position estimates along the x and y axes of the image.

### Correlation with ground truth

Pearson r between mean position estimates (with standard error of the mean as a measure of variance) by humans and ground truth (scaled to the dimensions of the images that humans saw) was computed for each condition and axis (x and y). Confidence intervals were estimated using Fisher transformation of the raw r value and plotted against their ceiling *c*

$$c = \sqrt{r_a * r_b}$$

Where $r_a$ and $r_b$ are split-half reliabilities of the two sets of position estimates being correlated.

### Parametric Statistics

Distribution of the means was assessed using the Shapiro-Wilk test from Scipy (1.11.2). If the data violated the assumptions of normality, differences between position estimates following different adapter conditions were assessed using the Wilcoxon Signed Rank test from Scipy (1.11.2). Else, differences were assessed using a two-tailed related samples t-test (Scipy 1.11.2).

### Mean Deltas

For both x and y position estimates, the difference between estimates before and after adaptation to rightward motion (rightward adapter $\Delta P = P_{adapt\ right} - P_{no\ adapt}$), before and after adaptation to leftward motion (leftward adapter $\Delta P = P_{adapt\ left} - P_{no\ adapt}$), as well as following adaptation to leftward and rightward motion ($\Delta P = P_{adapt\ left} - P_{adapt\ right}$), was computed for each image. The mean difference across all 40 presented images was termed as the mean delta, with standard error of the mean as a quantification of the variance. We tested whether mean deltas were significantly different from zero using the single sample t-test from Scipy (1.11.2).

## Neural metrics from Macaque IT responses

### Neural site inclusion criteria:

We detected the multiunit spikes after the raw voltage data were collected. A multiunit spike event was defined as the threshold crossing when voltage (falling edge) deviated by less than three times the standard deviation of the raw voltage values. Our array placements allowed us to sample neural sites from different parts of IT, along the posterior to anterior axis. However, for all the analyses, we did not consider the specific spatial location of the site, and treated each site as a random sample from a pooled IT population.

For our analyses, we only included neural recording sites that exhibited an overall significant visual drive, an image rank order response reliability greater than 0.7. Given that most of our neural metrics are corrected for the estimated noise at each neural site, the criterion for selecting neural sites is not critical, and it was primarily used to reduce computation time by eliminating noisy recordings.

To assess the reliability of individual neural sites, we computed a split-half internal consistency metric across stimulus repetitions. This measure quantifies how stable each site's response pattern is across random subsets of trials, providing an inclusion criterion for downstream analyses (e.g., only sites exceeding a reliability threshold of 0.7 were used for model–neural comparisons).

For each neural site, we repeatedly divided the available trials into two random, non-overlapping halves. Within each split, we computed the mean response of the neuron across all repetitions for each image. The correlation between these two split means was taken as the raw split-half reliability. To obtain a reliability estimate corrected for finite sampling, we applied the Spearman–Brown correction, which compensates for the halving of the data. This procedure was repeated across 100 random half-splits, and the final reliability for each site was defined as the average of these corrected correlations.

### Split-Half Reliability

Split-half reliability across 70-170 ms after image onset was used for filtering reliable neurons to predict object position from HVM640 and MAE40. For each neuron, average spike counts for the time bin between 70 ms and 170 ms after image onset were randomly split into two groups of equal sizes across trials. Responses to each image are averaged across trials and the two groups were correlated to each other (Pearson r, Scipy 1.11.2). The correlation was then corrected using the Spearman-Brown prediction formula, where the corrected correlation

$$r_{corrected} = \frac{2r}{r+1}$$

This process was repeated 20 times. The average of the split-half reliabilities across the 20 repetitions for each neuron was taken as that neuron's split-half reliability, with the standard deviation across repetitions computed to estimate the variance. Only units with split-half reliability $r_{corrected} > 0.2$ were used for subsequent analyses.

Correlation between neural responses before and after motion adaptation

To test unit-level stability across paradigms and sensitivity to motion adaptation, we correlated each neuronal unit's response to all images in the MAE40 set before and after adaptation to left and right motion. We first selected units whose responses had a split-half reliability $r_{corrected} > 0.2$ across 70-170 ms following stationary image presentation. For each unit, we computed the average spike count for this time segment, for all images, across a random subsample of half the trials, before and after motion adaptation. The resulting response vectors, one representing the response of one unit to each image before motion adaptation and the other representing the response of the same unit to the same images after motion adaptation, were correlated to each other (Pearson r, Scipy 1.11.2). The correlation was then corrected by its ceiling, computed in the same manner as for the correlation between human object position estimates and ground truth. This process was repeated 20 times for each unit and each direction of motion adaptation. For each neuronal unit and direction of motion adaptation, the mean correlation across 20 repetitions was plotted.

Centered kernel alignment

Before testing for systematic biases in object position decodes following motion adaptation, we examined the effect of motion adaptation on neural representations of stationary images using centered kernel alignment (CKA), as described in Kornblith et al., 2019. As always, we selected responses of units with split-half reliability $r_{corrected} > 0.2$ across 70-170 ms following stationary image presentation, z-scored (Scipy 1.11.2) responses across images and computed CKA between responses before and after adaptation to leftward or rightward motion. This process was bootstrapped 50 times across subsamples of images. To estimate the ceiling, we computed the CKA between unit responses, across 4 subsampled trials, before motion adaptation. This estimate was bootstrapped 10 times.

Predicting object position

Prior to predicting object position, responses were z-scored (Scipy 1.11.2) across the unit dimension and averaged across the trial dimension.

To test for object position information in macaque IT cortex, we decoded x- and y-object positions from model responses to MAE40 images. Responses of all model units to a training set of 30 images for MAE40 were used to train a linear decoder (L2-regularized, scikit-learn 1.3.0 Ridge) to predict the x- and y-positions of the object in the image. The reliability of each decoder was computed in the same manner as split-half reliability. Decoders were then tested on held-out images, and k-fold cross validation (k=4) produced predictions for all images. Fold-indexing was shuffled 100 times to estimate variability of prediction accuracy. The penalty weight for each for each split was chosen separately with cross-validation by sub-splitting the training data (scikit-learn 1.3.0, RidgeCV, α = [0.01, 0.1, 1, 10, 100]).

To examine the effects of motion adaptation on object position information, we used these decoders to predict object position from responses to MAE40 image presentation following adaptation to either rightward or leftward motion. Each decoder only predicted x- and y-positions for images that were held out during training.

### Reliability of Neural Predictions

We estimated the reliability of position predictions from macaque IT responses before and after motion adaptation by computing the correlation between predictions from split-halfs of neural data across trials.

To test the reliability of predictions before motion adaptation, we randomly split average spike counts for the time bin between 70 ms and 170 ms after test image onset (before adapter presentation) into two groups of equal sizes across trials. We then decoded x- and y-object positions from neural responses to MAE40 images, average across each random half of trials. Responses of neuronal units to a training set of 30 images for MAE40 were used to train a linear decoder (L2-regularized, scikit-learn 1.3.0 Ridge) to predict the x- and y-positions of the object in the image. Decoders were then tested on held-out images, and k-fold cross validation (k=4) produced predictions for all images. Fold-indexing was shuffled 100 times to estimate variability of prediction accuracy. The penalty weight for each for each split was chosen separately with cross-validation by sub-splitting the training data (scikit-learn 1.3.0, RidgeCV, α = [0.01, 0.1, 1, 10, 100]). Predictions for each image are averaged across shuffle iterations, and the two sets of predictions were correlated to each other (Pearson r, Scipy 1.11.2). The correlation was then corrected using the Spearman-Brown prediction formula and the process was bootstrapped 20 times. The average of the split-half reliabilities across the 20 repetitions taken as the prediction split-half reliability, with the standard deviation across repetitions computed to estimate the variance.

To test the reliability of predictions after motion adaptation, we randomly split average spike counts for the time bin between 70 ms and 170 ms after test image onset (which was preceded by adaptation to leftward or rightward motion) into two groups of equal sizes across trials. We used decoders trained on all reliable neural data to predict object positions in MAE40 images following motion adaptation. Each decoder only predicted x- and y-positions for images that were held out during training. The two sets of predictions for each direction of motion adaptation were correlated to each other (Pearson r, Scipy 1.11.2). The correlation was then corrected using the Spearman-Brown prediction formula and the process was bootstrapped 20 times. The average of the split-half reliabilities across the 20 repetitions taken as the prediction split-half reliability, with the standard deviation across repetitions computed to estimate the variance.

### Correlation with ground truth and parametric statistics

Pearson r between x- and y- position predictions from macaque IT neurons and ground truth was computed separately for each random shuffle of fold-indexing. The correlation $r$ for each shuffle was corrected by the decoder split-half reliability, such that

$$r_{corrected} = \frac{r}{\sqrt{s}}$$

where $s$ is the mean split-half reliability of the contributing decoders. The average correlation for predictions was then computed, with standard deviation across all correlation values as a measure of variability.

The same tests and methods for computing mean ΔP were used for position predictions from macaque IT neurons as for behavioural estimates. To understand the relationship between ΔP and the amount of neural data (number of neural units) used to predict object position, we randomly subsampled from our neural units and followed the previously described procedure to predict the positions of objects in MAE40 images before and after adaptation to rightward and leftward motion. We tested subsample sizes from 10 to 110 neural units, in intervals of 10. Random subsampling was repeated 20 times for each subsample size. Standard deviation across these repetitions provided a measure of variance in position predictions with the amount of neural data used.

### Estimating response decay

Similar to Vinken et al., (2020) we used an exponential decay function to quantify the amount of response decay in IT neurons during motion adaptation. As always, we selected only units with split-half reliability $r_{corrected}$ > *0.2*, 70-170 ms following test-image presentation. For each of these units, we used a nonlinear least-squares procedure (Scipy 1.11.2 optimize.curve_fit) to estimate the parameters of an exponential decay function that would best fit the first 1000 ms of neuronal response, averaged across trials, to motion adaptation. Specifically, we estimated parameters *C* and $\tau$ for a function

$$y = Ce^{\left(\frac{-t}{\tau}\right)}$$

that produces responses y closest to those of the neuron at time *t*. This was done separately for each unit's response to right and left motion adaptation. To assess the quality of the exponential decay fits, we computed the coefficient of determination ($R^2$; scikit-learn 1.3.0 metrics.r2_score) between each exponential decay function and corresponding unit response. Only units whose response to left or right motion adaptation was approximated by an exponential decay function with a goodness of fit of $R^2$ > *0.3* were included in the resulting $\tau$ parameter distribution.

## Responses from deep ANN units

### Preprocessing

For all static ANNs, except ViT-L32, extracted features had the shape [*N, C, H, W*] where *N* was the number of images, *C* the number of channels, and [*H, W*] the spatial dimensions of the convolutional layer. Features were reshaped into a matrix of [*N, (C\*H\*W)*] for subsequent analysis. In the case of ViT-L32, extracted features had the shape [*N*, 50, 1024], with 50 patches and 1024 being the size of the linear layer. Before reshaping, the element of the patch dimension corresponding to the classification token was removed. As a result, the reshaped ViT-L32 features were [*N*, 50176].

For SlowFast, there was an additional dimension *T*, denoting convolution over multiple sets of sampled frames (i.e., model time). This dimension was preserved and features at each time point were analyzed separately.

For ConvRNN, features out of each video were of shape [*T, H, W, C*] where *T*=238, *H*=*W*=4, *C*=512, which have been then linearized to [*T, (H\*W\*C)*] = [238, 8192].

Features were then z-scored (Scipy 1.11.2) across the unit dimension and min-max scaling (Numpy 1.26.4) was applied across the image dimension.

### Predicting object position

To test for object position information in our deep ANNs, we decoded x- and y-object positions from model responses to HVM640 and MAE40 images separately. Responses of all model units to a training set of 576 images for HVM640 and 30 images for MAE40 were used to train a linear decoder (L2-regularized, scikit-learn 1.3.0 Ridge) to predict the x- and y-positions of the object in the image. Decoders were then tested on held-out images, and *k*-fold cross validation *k* = 10 for HVM640 and *k* = 4 for MAE40) produced predictions for all images. Fold-indexing was shuffled 100 times to estimate variability of prediction accuracy. The penalty weight for each for each split was chosen separately with cross-validation by sub-splitting the training data (scikit-learn 1.3.0, RidgeCV, α = [0.01, 0.1, 1, 10, 100]).

#### SlowFast

X- and y-positions were decoded across the 8 sampled time-points of the Slow stream independently. To examine the effects of motion adaptation on object position information, we used these decoders to predict object position from responses to MAE40 image presentation following adaptation to either rightward or leftward motion. Decoders trained on each sampling time point (of 8 Slow sampling time points) were tested on corresponding time points in videos that included motion adaptation. Each decoder only predicted x- and y-positions for videos that were held out during training.

#### ConvRNN

X- and y-positions were decoded from features corresponding to model unit responses averaged across a window of 100-200 ms following MAE40 image presentation. To examine the effects of motion adaptation on object position information, we used these decoders to predict object position from responses to MAE40 image presentation following adaptation to either rightward or leftward motion. Each decoder only predicted x- and y-positions for images that were held out during training.

### Correlation with ground truth and parametric statistics

Pearson r between x- and y- position predictions from deep artificial neural networks and ground truth was computed separately for each random shuffle of fold-indexing. The average correlation for predictions from a given model was then computed, with standard deviation across all correlation values as a measure of variability.

The same tests and methods for computing mean deltas were used for position predictions from ANNs as for behavioural estimates and macaque IT neural responses.

# Supplementary Materials

The macaque IT cortex but not current deep networks encodes object position in perceptually aligned coordinates.

*Elizaveta Yakubovskaya, Hamidreza Ramezanpour, Matteo Dunnhofer, and Kohitij Kar*

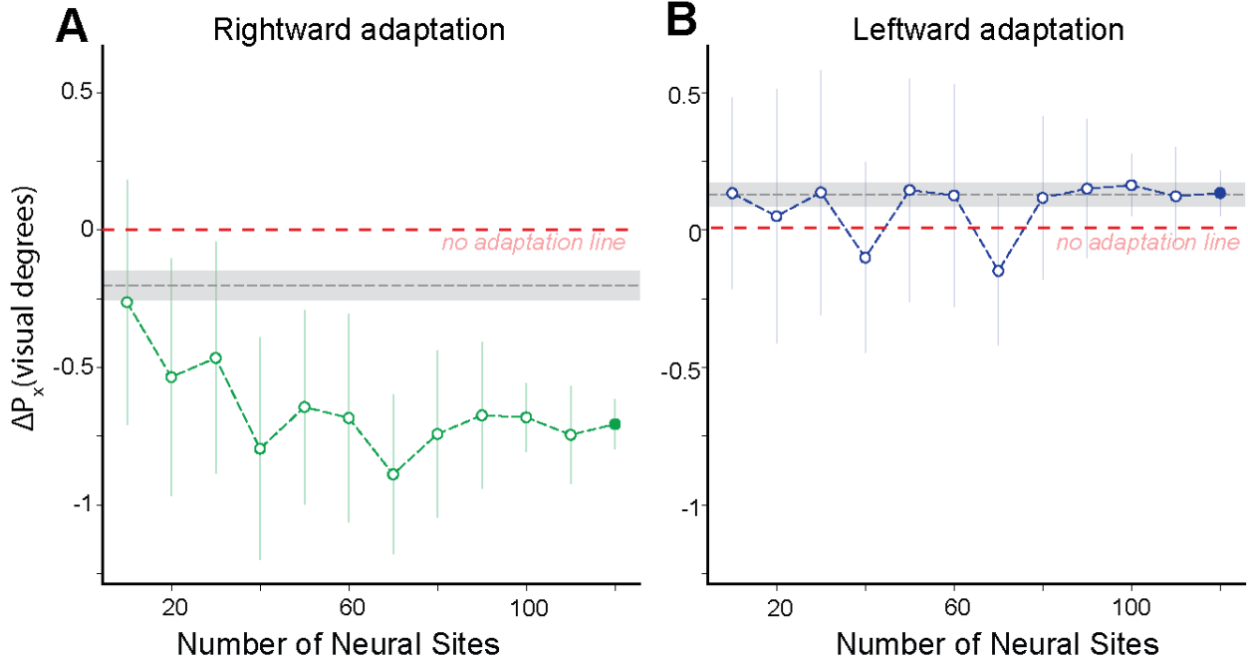

**Supplementary Figure 1. Motion adaptation induces biases in object position predictions that align with those reported by humans.** Group-level biases following **(A)** rightward motion adaptation ($\Delta P_x = P_{x,\text{ adapt right}} - P_{x,\text{ no adapt}}$) and **(B)** leftward motion adaptation ($\Delta P_x = P_{x,\text{ adapt left}} - P_{x,\text{ no adapt}}$) in x-position predictions as a function of the number of neural units contributing to predictions. The final, colored point denotes the same statistic from the whole population of neurons. Error bars denote standard deviation across subsamples. The grey dashed line indicates group-level biases computed from human position estimates and the shaded region denotes standard error of the mean across trials. Biases in x-position predictions from macaque IT have the same directionality (i.e., opposite the adapter direction) as those in human position estimates.

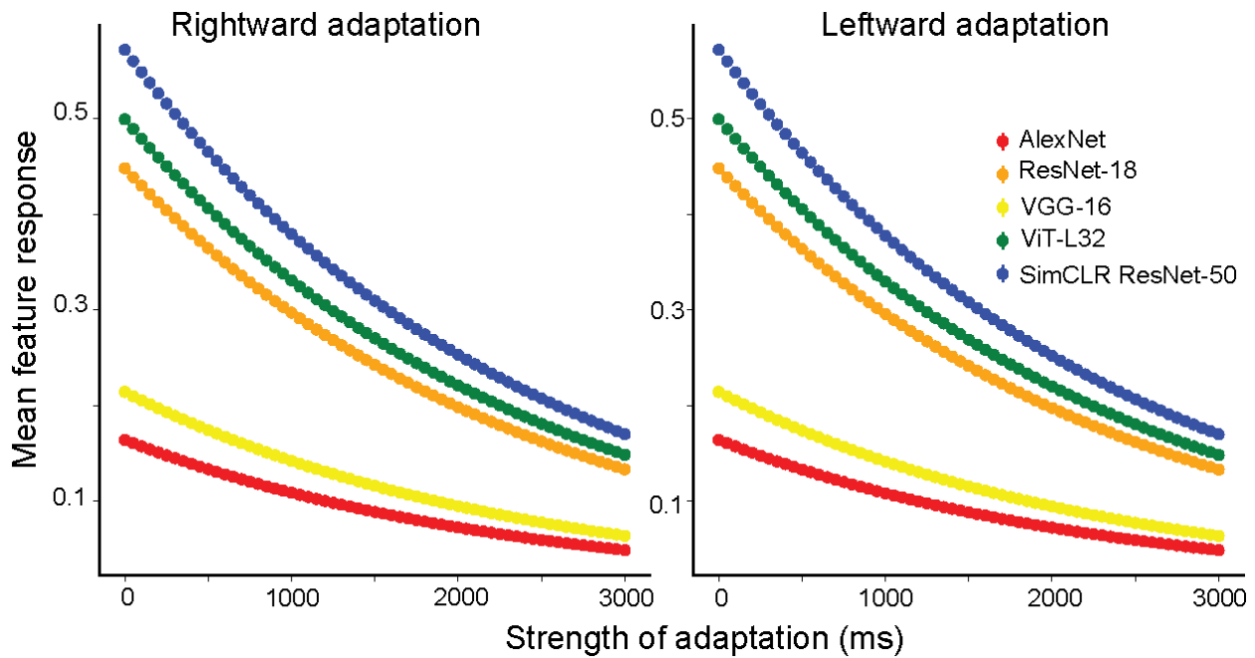

**Supplementary Figure 2.** Mean model unit responses during simulated adaptation to rightward (left plot) and leftward (right plot) motion, demonstrating decay trajectories across ANN units.

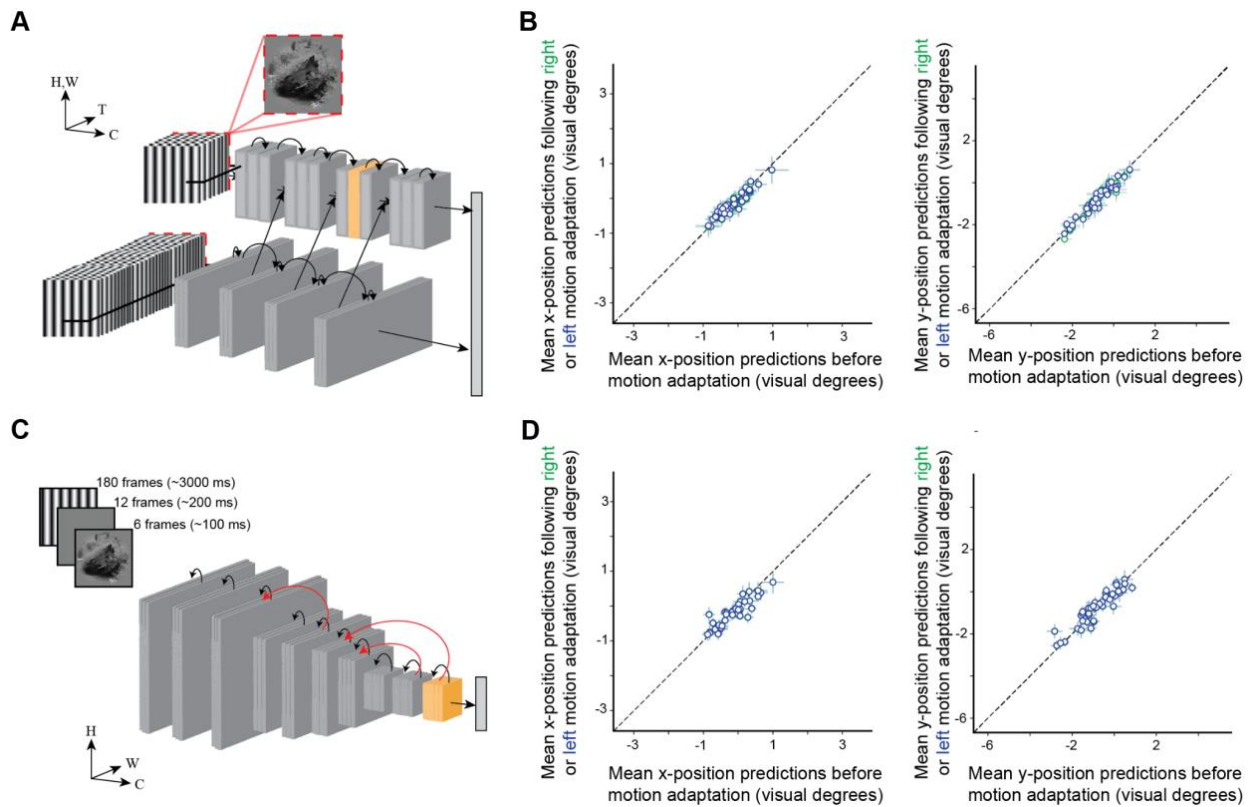

**Supplementary Figure 3. Temporal and recurrent architectures fail to reproduce IT-like adaptation biases. (A)** Schematic of the SlowFast architecture with adapter and test sequences. Features were extracted from IT-aligned layers in the slow pathway. **(B)** Scatterplots of predicted positions following rightward versus leftward adaptation for x- (middle) and y- (right) dimensions. Predictions remained tightly aligned to the unity line, indicating no systematic positional bias. **(C)** Schematic of ConvRNN architecture with temporally structured input (180 adapter frames, 200 ms gray interval, 100 ms test image). **(D)** Scatterplots of position predictions following rightward versus leftward adaptation for x- (middle) and y- (right) dimensions. Predictions again aligned with the unity line, showing no effect of adaptation. Across both architectures, temporal convolution and recurrence failed to generate the direction-opponent position shifts observed in macaque IT and human observers.